\documentclass[]{jss}

\usepackage[utf8]{inputenc}

\author{
Zihuan Qiao\\Boston University \And Daniel Sussman\\Boston University
}
\title{\pkg{iGraphMatch}: an \proglang{R} Package for the Analysis of Graph Matching}

\Plainauthor{Zihuan Qiao, Daniel Sussman}
\Plaintitle{iGraphMatch: an R Package for the Analysis of Graph Matching}
\Shorttitle{\pkg{iGraphMatch} \proglang{R} Package for Graph Matching}

\Abstract{
\pkg{iGraphMatch} is an \proglang{R} package for finding corresponding vertices between two graphs, also known as graph matching. The package implements three categories of prevalent graph matching algorithms including relaxation-based, percolation-based, and spectral-based, which are applicable to matching graphs under general settings: weighted directed graphs of different order and graphs of multiple layers. We provide versatile options to incorporate prior information in the form of seeds with or without noise and similarity scores. In addition, \pkg{iGraphMatch} provides functions to summarize the graph matching results in terms of several evaluation measures and visualize the matching performance. Finally, the package enables users to sample correlated random graph pairs from classic random graph models to generate data for simulations. This paper illustrates the practical applications of the package to the analysis of graph matching by detailed examples using real data from communication, neuron, and transportation networks.
}

\Keywords{graph matching algorithms, \proglang{S}4, \proglang{R}}
\Plainkeywords{graph matching algorithms, S4, R}

\usepackage{subfig}

\usepackage{thumbpdf,lmodern}
\usepackage{framed}
\usepackage[utf8]{inputenc}
\usepackage{graphicx,algpseudocode}
\usepackage{geometry, amssymb, amsmath, amsthm}
\usepackage{bbm}
\usepackage{enumitem}
\usepackage{chngpage}
\usepackage{breqn}
\usepackage{algorithm}
\usepackage[algo2e]{algorithm2e}
\usepackage[]{algpseudocode}
\usepackage{amsfonts}
\usepackage{dsfont}
\usepackage{multirow}
\usepackage{longtable}
\usepackage{tabularx,booktabs}

\newtheorem{defn}{Definition}

\begin{document}

\hypertarget{sec:intro}{%
\section{Introduction}\label{sec:intro}}

The graph matching (GM) problem seeks to find an alignment between the vertex sets of graphs that best preserves common structure across graphs.
This is often posed as minimizing edge disagreements of two graphs over all alignments.
Formally, given \(A\) and \(B\), two adjacency matrices corresponding to two graphs \(G_1=(V_1, E_1)\) and \(G_2=(V_2, E_2)\), the goal is to find
\begin{align*}
\mathop{\mathrm{argmin}}_{P\in\Pi}\lVert A-PBP^\top \rVert_F^2
\end{align*}
where $\Pi$ is the set of all permutation matrices.
GM has wide applications in diverse fields, such as pattern recognition \citep{PatternRec1, PatternRec2, PatternRec3}, machine learning \citep{ML1, ML2}, bioinformatics \citep{bio3, bio2}, neuroscience \citep{neuro}, social network analysis \citep{SocialNetwork}, and knowledge graph queries \citep{Hu2018-hd}.
More generally, the problem of discovering some true latent alignment between two networks can often be posed as variations on the above problem by adjusting the objective function for the setting.

The well-known graph isomorphism problem is a special case of GM problem when there exists a bijection between the nodes of two graphs which exactly preserves the edge structure.
In terms of computational complexity, GM is equivalent to the NP-hard quadratic assignment problem, which is considered a very challenging problem where few theoretical guarantees exist, even in special cases \citep{QAP}.
For certain problems where the graphs are nearly isomorphic, polynomial-time algorithms do exist \citep{friendly, Umeyama} but these methods frequently break down for more challenging instances.

This paper presents the detailed functionality of the \pkg{iGraphMatch} \proglang{R} package which serves as a practical tool for the use of prevalent graph matching methodologies.
These algorithms utilize either the spectral embedding of vertices \citep{Umeyama}, or relaxations of the objective function \citep{PATH, FAQ}, or apply ideas from percolation theory \citep{Percolation, ExpandWhenStuck}.
The \pkg{iGraphMatch} package provides versatile options of working with graphs in the form of matrices, igraph objects or lists of either, and matching graphs under a generalized setting: weighted, directed, graphs of a different order, and multilayer graphs.

In addition, the \pkg{iGraphMatch} package incorporates prior information: seeds and similarities for all the implemented algorithms.
Seeds, or anchors, refer to partial knowledge of the alignment of two graphs.
In practice, seeds can be users with the same name and location across different social networks or pairs of genes with the same DNA sequences.
Some algorithms like the percolation algorithm \citep{Percolation, ExpandWhenStuck} which matches two graphs by propagating matching information to neighboring pairs require seeds to kick off.
All algorithms improve substantially by incorporating seeds and can achieve accurate matching in polynomial time \citep{SGM}.
Similarity scores are another commonly used prior which measures the similarity between pairs of nodes across the graphs.
In the bioinformatics area, BLAST similarity score is an example of similarity scores that plays an important role in aligning two PPI networks \citep{IsoRank}.
Similarity scores are usually generated from nodal covariates that are observed in both networks \citep{BLAST, SimilarityScore}.

While under many scenarios the availability of exact partial matches, or hard seeding, is not realistic and expensive, the package also enables utilizing noisy prior information.
Similarity scores incorporate uncertainty by assigning the pair of nodes with higher similarity scores a bigger chance to match.
Seeds with uncertainty and even error can still be handled by self-correcting graph matching algorithms like the \textsc{Frank-Wolfe} algorithm initialized at the noisy partial matching, called soft seeding.
\cite{soft_seeding} showed that the \textsc{Frank-Wolfe} algorithm with soft seeding scheme converges quickly to the true alignment under the correlated {Erd{\H o}s-R{\'e}nyi} model with high probability,
provided sufficient correct information is available.

Although there exist some open source software and packages containing graph matching functionality, \pkg{iGraphMatch} package provides a centralized repository for common graph matching methodologies with flexibility, tools for developing graph matching problem methodology, as well as metrics for evaluating and tools for visualizing matching performance.
Among the alternative GM packages, the most relevant ones include
the \pkg{igraph} \citep{igraph} package which focuses on descriptive network analysis and graph visualization based on igraph objects and provides a single graph matching algorithm,
the \pkg{GraphM} \citep{PATH} package which implements several GM algorithms proposed between 1999 and 2009 in \proglang{C}, and the
\pkg{Corbi} \citep{Corbi} \proglang{R} package which is particularly designed for studies in bioinformatics and \pkg{SpecMatch} \citep{SpecMatch} which only involves implementations of spectral embedding based GM algorithms and written in \proglang{C/C++}.
None of these packages provide the breadth of state-of-the-art tools, flexibility, and ease-of-use provided by the \pkg{iGraphMatch} package.

The rest of this paper is organized as follows.
Section \ref{sec:background} describes the theoretical representations of the implemented GM algorithms, correlated random graph models and evaluation metrics.
Section \ref{sec:usage} discusses the functionality and usage of \proglang{R} functions in the package, illustrated on the synthetic correlated graph pairs.
Section \ref{sec:example} presents more complex examples on real data with several functions involved in the analysis and Section \ref{sec:conclusion} gives guidelines on using different GM algorithms under different circumstances and concludes the paper.

\hypertarget{sec:background}{%
\section{Graph matching background}\label{sec:background}}

In this section, we give background on graph matching and related problems followed by descriptions of the principal algorithms implemented in \pkg{iGraphMatch}.
For simplicity, we state all the algorithms in the context of matching undirected, unweighted graphs with the same cardinality.
All algorithms can also be directly applied to directed and weighted graphs.
In the second subsection, we discuss the techniques for matching graphs with a different number of vertices along with other extensions.
To conclude the section, we introduce the statistical models for correlated networks and discuss measures for the goodness of matching.

For the remainder of this paper we use the following notations.
Let \(G_1=(V_1,E_1)\) and \(G_2=(V_2,E_2)\) denote two graphs with \(n\) vertices.
Let \(A\) and \(B\) be their corresponding binary symmetric adjacency matrices.
In the setting of seeded graph matching, suppose without loss of generality, the first \(s\) pairs of nodes are seeds for simplicity.
In \pkg{iGraphMatch}, much more flexible seed specifications are possible, which will be illustrated in examples for usage of the package in Section \ref{sec:usage}.
Accordingly, let \(A\) and \(B\) be partitioned as:
\begin{equation} \label{eq:seed_blocks}
A=
\begin{bmatrix}
A_{11} & A_{21}^\top \\
A_{21} & A_{22}
\end{bmatrix}\text{ and }
B=
\begin{bmatrix}
B_{11} & B_{21}^\top \\
B_{21} & B_{22}
\end{bmatrix}
\end{equation}
where \(A_{11}, B_{11}\in\{0, 1\}^{s\times s}\) denote seed-to-seed adjacencies, \(A_{21}, B_{21}\in\{0, 1\}^{(n-s)\times s}\) denote nonseed-to-seed adjacencies and \(A_{22}, B_{22}\in\{0, 1\}^{(n-s)\times (n-s)}\) denote nonseed-to-nonseed adjacencies.
Let \(S\) be an \(n\)-by-\(n\) real-valued matrix of similarity scores.
Let \(\Pi\) be the set of all permutation matrices and \(\mathcal{D}\) be the set of all doubly stochastic matrices.

\hypertarget{assignment-problems}{%
\subsubsection{Assignment problems}\label{assignment-problems}}

Matching or assignment problems are core problems in combinatorial optimization and appear in numerous fields \citep{AP}.
As we illustrate in Equation\textasciitilde{}\ref{eq:ob_func}, a general version of the graph matching problem is equivalent to the quadratic assignment problem (QAP).
Similarly, QAP is related to the linear assignment problem (LAP) which also plays a role in GM.
The LAP asks how to assign \(n\) items (eg. workers or nodes in \(G_1\)) to \(n\) other items (eg. tasks or nodes in \(G_2\)) with minimum cost.
Let \(C\) denote an \({n\times n}\) cost matrix, where \(C_{ij}\) denotes the cost of matching \(i\) to \(j\), then the LAP is to find
\begin{equation} \label{eq:LAP}
\begin{aligned}
& \mathop{\mathrm{argmin}}_{P\in \Pi}
& & \mathrm{trace}(C^\top P)
\end{aligned}
\end{equation}
LAP is solvable in \(O(n^3)\) time and there are numerous exact and approximate methods for both general \citep{lap_solver, Hungarian} and special cases, such as sparse cost matrices \citep{Volgenant1996-o}.

The statement of QAP resembles LAP, except that the cost function is expressed as a quadratic function.
Given two \(n\text{-by-}n\) matrices \(A\) and \(B\) which can represent flows between facilities and the distance between locations respectively, or the adjacency matrices of two unaligned graphs, the objective function for QAP is:
\begin{equation}\label{eq:qap}
\mathop{\mathrm{argmin}}_{P\in \Pi}    \mathrm{trace}(APBP^\top).
\end{equation}
This problem is NP-hard \citep{QAP} leading to a core challenge for any graph matching approach.

As will be illustrated in the rest of the section, some matching algorithms reduce the graph matching problem to solving a LAP.
For these algorithms, we include similarity scores \(S\) by adding an additional term \(\mathrm{trace}(S^\top P)\) to the reduced objective function.

\hypertarget{subsec:gm}{%
\subsection{Graph matching algorithms}\label{subsec:gm}}

In the \pkg{iGraphMatch} package, we implement three types of prevalent GM algorithms.
The first group uses relaxations of the objective function, including convex, concave, and indefinite relaxations.
The second group consists of algorithms that apply ideas from percolation theory, where matching information is spread from an initial set of matched nodes.
The last group is based on the spectral embedding of vertices.

\hypertarget{relaxation-based-algorithms}{%
\subsubsection{Relaxation-based algorithms}\label{relaxation-based-algorithms}}

These approaches relax the constraint that \(P\) is a permutation matrix to require only that \(P\) is doubly stochastic, optimizing over \(\mathcal{D}\), the convex hull of \(\Pi\).
When \(P\) is a permutation matrix
\begin{equation}
\label{eq:ob_func}
    \lVert A-PBP^\top \rVert_F^2 = \lVert AP-PB \rVert_F^2 = \lVert A \rVert_F^2 + \lVert B \rVert_F^2 - 2\cdot \mathrm{trace}APBP^\top.
\end{equation}
However, these equalities do not hold for all \(P\in \mathcal{D}\), leading to different relaxations.

The second term of Equation\textasciitilde{}\ref{eq:ob_func} is a convex function and optimizing it over \(P\in\mathcal{D}\) gives the convex relaxation, where the gradient at \(P\) to the convex relaxed objective function is \(-4 APB + 2A^\top AP + 2PBB^\top\).
The last equality in Equation\textasciitilde{}\ref{eq:ob_func} shows that minimizing edge disagreements is equivalent to maximizing the number of edge agreements, \(\mathrm{trace} APBP^\top\), a QAP.
Optimizing the indefinite function over \(\mathcal{D}\) gives the indefinite relaxation with gradient \(-2APB\) \citep{FAQ}.

\begin{longtable}[h]{lccccc}
\toprule\addlinespace
Relaxation           & \begin{tabular}[c]{@{}c@{}}Objective \\ function\end{tabular}                                     & Domain        & \begin{tabular}[c]{@{}c@{}}GM\\ algorithm\end{tabular} & \begin{tabular}[c]{@{}c@{}}Optimization \\ guarantee\end{tabular} & \begin{tabular}[c]{@{}c@{}}Optimum \\ form\end{tabular}
\\\addlinespace
\midrule\endhead
None        & $\lVert A-PBP^T \rVert_F^2$    & $\Pi$                & NA   &     & $\Pi$
\\\addlinespace
Indefinite  & $trBDAD^T$                     & $\mathcal{D}$   & FW  & Local  & $\mathcal{D}$ (often $\Pi$)
\\\addlinespace
Convex      & $\lVert AD-DB \rVert_F^2$      & $\mathcal{D}$   & FW, PATH & Global & $\mathcal{D}$
\\\addlinespace
Concave  & \begin{tabular}[c]{@{}c@{}}$-tr(\Delta D)-$\\ $2tr(L_1^TDL_2D^T)$\end{tabular} & $\mathcal{D}$ & PATH  & Local & $\Pi$
\\\addlinespace
\bottomrule
\addlinespace
\caption{Summary of relaxation methods for graph matching problem}
\label{tab:relaxation}
\end{longtable}

Generally, the convex relaxation leads to a solution that is not guaranteed to be near the solution to the original GM.
However, \cite{friendly} introduced the class of ``friendly'' graphs based on the spectral properties of the adjacency matrices to characterize the applicability of the convex relaxation.
Matching two friendly graphs by using the convex relaxation is guaranteed to find the exact solution to the GM problem.
Unfortunately, this class is quite limiting and does not hold for most statistical models or real-world examples.

Another relaxation is the concave relaxation used in the PATH algorithm \citep{PATH}.
The concave relaxation uses the Laplacian matrix defined as \(L=D-A\), where \(D\) is the diagonal degree matrix with diagonal entries \(D_{ii}=\sum_{i=1}^N A_{ij}\).
Assume \(L_i\) and \(D_i\), \(i=1,2\), are the Laplacian matrices and degree matrices for \(G_1\) and \(G_2\) respectively, then we can rewrite the objective function as
\begin{equation} \label{eq:concave}
\begin{split}
    \lVert A-PBP^\top\rVert_F^2 & = \lVert AP-PB \rVert_F^2\\
    & = \lVert (D_1P-PD_2)-(L_1P-PL_2)\rVert_F^2\\
    & = -\mathrm{trace}(\Delta P)+\mathrm{trace}(L_1^2)+\mathrm{trace}(L_2^2)-2\mathrm{trace}(L_1^\top P L_2 P^\top),
\end{split}
\end{equation}
where the matrix \(\Delta_{ij}=(D_{2_{jj}}-D_{1_{ii}})^2\).
Dropping the terms not dependent on \(P\) in Equation\textasciitilde{}\ref{eq:concave}, we obtain the concave function \(-\mathrm{trace}(\Delta P)-2\mathrm{trace}(L_1^\top P L_2 P^\top)\) on \(\mathcal{D}\).

A summary of the different relaxations is provided in Table\textasciitilde{}\ref{tab:relaxation}.
Relaxing the discrete problem to a continuous problem breaks the equivalence to the original formulation of the edge disagreement and enables employing algorithms based on gradient descent.

\hypertarget{methodology}{%
\paragraph{\texorpdfstring{\textsc{Frank Wolfe} methodology}{ methodology}}\label{methodology}}

\cite{FAQ} introduced an algorithm for the relaxed graph matching problem, with each iteration computable in polynomial time, that can find local optima for the relaxations above.
The \textsc{Frank-Wolfe (FW)} \citep{FW} methodology is an iterative gradient ascent approach composed of two steps.
The first step finds an ascent direction that maximizes the gradient ascent.
In this case the ascent direction is a permutation matrix which is a vertex of the polytope of doubly stochastic matrices.
For the convex, indefinite, and concave relaxations, this corresponds to a LAP with the gradient as the cost function.
The second step performs a line search along the ascent direction to optimize the relaxed objective function.
As the objectives are all quadratic, this line search simply requires optimizing a single-variable quadratic function along a line segment.
After the iterative algorithm converges, the final step of the procedure is to project the doubly stochastic matrix back to the set of permutation matrices, which is also a LAP.

The various relaxed forms can all serve as the objective function \(f(\cdot)\) in the \textsc{FW} Methodology, but in all cases a matrix \(D^0\in \mathcal{D}\) must be chosen to initialize the procedure.
For the convex relaxation, the \textsc{FW} methodology is guaranteed to converge to the global optimum regardless of the \(D^0\).
On the other hand, the \textsc{FW} algorithm for the indefinite relaxation is not guaranteed to find a global optimum so the initialization is critical.

In many instances, the optimal solution to the convex relaxation lies in the interior of \(\mathcal{D}\).
This can lead to inaccurate solutions after the last projection step.
The local optima for the indefinite relaxation are often at extreme points of \(\mathcal{D}\), meaning the final projection often does nothing.

The default initialization for the indefinite problem is at the barycenter matrix, \(D^0 = \frac{1}{n}\mathbbm{1}\mathbbm{1}^\top\), but many other initialization procedures can be used.
These include randomized initializations, initializations based on similarity matrices, and initializing the indefinite relaxation at the interior point solution of the convex relaxation \citep{relax_paper}.
When prior information regarding a partial correspondence is known to be noisy, rather than incorporating this information as seeds, one can incorporate it as ``soft'' seeds which are used to generate the initialization \citep{soft_seeding}.

\begin{algorithm}
\caption{Frank-Wolfe Methodology}
\label{alg:FAQ}
\SetKwInOut{Input}{Input}
\SetKwInOut{Output}{Output}
\Input{$A,B$, doubly stochastic matrix $D^0$, tolerance $\epsilon$}
\Output{permutation matrix $P$}
Set $i=0$;\par
\While{$\Vert D^i-D^{i-1} \Vert^2_F\ge \epsilon$}{
    $P^i=$ $\mathop{\mathrm{argmin}}_{P\in\Pi}\mathrm{trace}\nabla f(D^i)^\top P$;\par
    $D^{i+1}=$ $\mathop{\mathrm{argmin}}_{D\in\mathcal{D}}f(D)$ over line segment from $D^i$ to $P^i$;\par
    $i=i+1$;\par
}
Project $D^{i}$ to the nearest $P$ by maximizing $\mathrm{trace}(P^\top D)$;\par
\end{algorithm}

When prior information is available in the form of seeds,
the seeded graph matching problem \citep{SGM} works on the objective function (\ref{eq:ob_func}) with the permutation matrix \(P^{n\times n}\) substituted by \(I_s\oplus P^{(n-s)\times (n-s)}\), the direct sum of an \(s\times s\) identity matrix and an \((n-s)\times (n-s)\) permutation matrix.
Employing the indefinite relaxed objective function incorporating seeds, we formulate the problem as finding
\begin{align*}
    \hat{P} 
    &= \mathop{\mathrm{argmax}}_{P\in\mathcal{D}} 2\cdot\mathrm{trace}P^\top A_{21}B_{21}^\top +\mathrm{trace}A_{22}PB_{22}P^\top
\end{align*}
where the gradient to the objective function is
\begin{equation} \label{eq:sgm_gradient}
  \nabla f(P)=2\cdot A_{21}B_{21}^\top +2\cdot A_{22}PB_{22}.
\end{equation}

In total, this uses the information between seeded nodes and nonseeded nodes and the nonseed-to-nonseed information.
Applying seeded graph matching to the convex relaxation and concave relaxation closely resembles the case of indefinite relaxation.

\hypertarget{algorithm}{%
\paragraph{\texorpdfstring{\textsc{PATH} algorithm}{ algorithm}}\label{algorithm}}

\cite{PATH} introduced a convex-concave programming approach to approximately solve the graph matching problem.
The concave relaxation has the same solution as the original graph matching problem.
The \textsc{PATH} algorithm finds a local optimum to the concave relaxation by considering convex combinations of the convex relaxation \(F_0(P)\) and the concave relaxation \(F_1(P)\) denoted by \(F_{\lambda} =(1 - \lambda) F_0 + \lambda F_1\).
Starting from the solution to the convex relaxation (\(\lambda=0\)) the algorithm iteratively performs gradient ascent using the FW methodology at \(F_\lambda\), increasing \(\lambda\) after each iteration, until \(\lambda = 1\).

\hypertarget{percolation-based-algorithms}{%
\subsubsection{Percolation-based algorithms}\label{percolation-based-algorithms}}

Under the \textsc{FW} methodology, all the nodes admit a correspondence but the (relaxed) matching correspondence evolves through iterations.
On the other hand, percolation approaches start with a set of seeds, adding one new match at each iteration.
The new matches are fixed and hence not updated in future iterations.

Each iteration expands the set of matched nodes by propagating the current matching information to neighbors.
The guiding intuition is that more matched neighbors are an indicator of a more plausible match, an intuition analogous to the gradient ascent approaches above.
We will present two algorithms in this category where the \textsc{ExpandWhenStuck} algorithm is an extension to the \textsc{Percolation} algorithm.

There are some distinctions about the inputs and outputs of percolation methods compared to the above relaxation methods.

\hypertarget{algorithm-1}{%
\paragraph{\texorpdfstring{\textsc{Percolation} Algorithm}{ Algorithm}}\label{algorithm-1}}

\cite{Percolation} provide a simple and fast approach to solve the graph matching problem by starting with a handful of seeds and propagating to the rest of the graphs.
At each iteration, the matching information up to the current iteration is encoded in a subpermutation matrix \(P\) where \(P_{ij}=1\) if \(i\) is matched to \(j\), and \(0\) otherwise.
The \textsc{percolation} algorithm searches for the most promising new match among the unmatched pairs through the mark matrix, \(M=APB\),
which is the gradient of the indefinite relaxation when extended to sub-doubly stochastic matrices.
When similarity scores are available, they are added to the mark matrix to combine topological structure and similarity scores.

Adopting analogous partitions on the adjacency matrices as in Equation\textasciitilde{}\ref{eq:seed_blocks}, we let \(A_{21}, B_{21}\) denote sub-matrix corresponding to potential adjacencies between unmatched and matched nodes.
Since all the candidates of matched pairs are permanently removed from consideration, we need only consider \(M'=A_{21}B_{21}^\top\), the sub-matrices of \(M\) corresponding to the unmatched nodes in both graphs.
As a result, the \textsc{percolation} algorithm only uses matched-to-unmatched information to generate new matches.

Moreover, the mark matrix \(M\) can also be interpreted as encoding the number of matched neighboring pairs for each pair of nodes \(i\in V_1\), \(j\in V_2\).
Suppose \(u, u'\in V_1\), \(v,v'\in V_2\), \([u,u']\in E_1\) and \([v,v']\in E_2\), then \((u',v')\) is a neighboring pair of \((u,v)\).
In each iteration, while there remain unmatched nodes with more than \(r\) matched neighboring pairs, the percolation algorithm matches the pair of nodes with the highest score \(M_{uv}\), and adds one mark to all the neighboring pairs of \((u,v)\).
Note that the algorithm may stop before all nodes are matched, leading to the return of a partial match.

There is only one tuning parameter in the \textsc{percolation} algorithm, the threshold \(r\) which controls a tradeoff between quantity of matches and quality of matches.
With a small threshold, the algorithm has a larger chance of matching wrong pairs.
If \(r\) is larger, then the algorithm might stop before matching many pairs \citep{ExpandWhenStuck}.

The \textsc{percolation} algorithm can be generalized to matching weighted graphs by making an adjustment to how we measure the matching information from the neighbors.
Since we prefer to match edges with smaller weight differences and higher absolute weights, we propose to adopt the following update formula for the score associated with each pair of nodes \((i,j)\):
\[M_{ij}=M_{ij} + \sum_{u\in N(i)}\sum_{v\in N(j)}1-\frac{|w_{iu}-w_{jv}|}{\max(|w_{iu}|, |w_{jv}|)}.\]
Thus, the score contributed by each neighboring pair of \((i,j)\) is a number in \([0,1]\).

\begin{algorithm}[H]
\SetAlgoLined
\SetKwInOut{Input}{Input}
\SetKwInOut{Output}{Output}
\Input{$A$, $B$, $s$ pairs of seeds, threshold $r$}
\Output{(sub-)permutation matrix $P$}
 Initialize the sub-permutation matrix $P$ incorporating seeds;\par
 Calculate the mark matrix $M=APB$;\par
 Set the rows and columns of $M$ corresponding to seeds to minus infinity;\par
 \While{$max(M)\ge r$}{
    $P_{ij}\leftarrow 1$, where $[i,j]$ is the index of $max(M)$;\par
    $M\leftarrow APB$;\par
    Set $i^{th}$ row and $j^{th}$ column of $M$ to minus infinity;\par
 }
\caption{Percolation Algorithm}
\label{alg:perco}
\end{algorithm}

\hypertarget{algorithm-2}{%
\paragraph{\texorpdfstring{\textsc{ExpandWhenStuck} Algorithm}{ Algorithm}}\label{algorithm-2}}

\cite{ExpandWhenStuck} extends the \textsc{percolation} algorithm to a version that can operate with a smaller number of seeds.
Without enough seeds, when there are no more unmatched pairs with a score higher or equal to the threshold \(r\), the \textsc{percolation} algorithm would stop even if there are still unmatched pairs.
\textsc{ExpandWhenStuck} uses all pairs of nodes with at least one matched neighboring pair, \(M_{ij}\geq 1\), as new seeds to restart the matching process by adding one mark to all of the new seeds' neighboring pairs, without updating the matched set.
If the updated mark matrix consists of new pairs with marks greater or equal to \(r\), then the percolation algorithm continues, leading to larger matched sets.

\hypertarget{spectral-based-algorithm}{%
\subsubsection{Spectral-based algorithm}\label{spectral-based-algorithm}}

Another class of graph matching algorithms uses the spectral properties of adjacency matrices.

\hypertarget{algorithm-3}{%
\paragraph{\texorpdfstring{\textsc{IsoRank} Algorithm}{ Algorithm}}\label{algorithm-3}}

\cite{IsoRank} propose the \textsc{IsoRank} algorithm that uses neighboring topology and similarity scores and exploits spectral properties of the solution.
The \textsc{IsoRank} algorithm is also based on the relaxation-based algorithms by encoding the topological structure of two graphs in \(ADB\), which is again proportional to the gradient of the indefinite relaxation.
However, the representations of each term of \(ADB\) are slightly different.
\(A\) and \(B\) are the column-wise normalized adjacency matrices and \(D\) is not necessarily a doubly stochastic matrix yet \(D_{ij}\) still indicates how promising it is to match \(i\in V_1\) to \(j\in V_2\).

Similar to the idea of \textsc{Percolation} algorithm, the intuition is that the impact of a pair of matched nodes is evenly distributed to all of their neighbors to propagate plausible matches.
This is achieved by solving the eigenvalue problem
\begin{align}\label{eq:Iso1}
    \mathrm{vec}(D)=(A\otimes B) \mathrm{vec}(D),
\end{align}
where \(\mathrm{vec}(D)\) denotes the vectorization of matrix \(D\), and the right hand side is equivalent to \(ADB\).
To combine network-topological structure and similarity scores in the objective function, the normalized similarity score \(E\) is added to the right hand side of Equation\textasciitilde{}\ref{eq:Iso1}, where \(E=S/\|S\|_1\), and \(\|\cdot\|_1\) denotes the L1 norm.

Note that when similarity score is not available as prior information, we can also construct a doubly stochastic similarity score matrix from seeds by taking \(I_{s\times s}\oplus \frac{1}{n-s}\mathbbm{1}\mathbbm{1}^\top_{(n-s)\times (n-s)}\).
To solve the eigenvalue problem in Equation\textasciitilde{}\ref{eq:Iso1}, we resort to the power method.
Finally, the global alignment is generated by a greedy algorithm or using the algorithms for solving the linear assignment problem (LAP).

\begin{algorithm}
\caption{IsoRank Algorithm}
\label{alg:IsoRank}
\SetKwInOut{Input}{Input}
\SetKwInOut{Output}{Output}
\Input{$A,B$, similarity scores $S$}
\Output{permutation matrix $P$}
Column-wise normalize the adjacency matrices $A$ and $B$;\par
Normalize similarity scores: $E=S/|S|_1$;\par
Initialize $D^0=E$ and tolerance $\epsilon$;\par
\While{$| D^i-D^{i-1}|\ge \epsilon$}{
    Calculate $D^{i+1}=AD^iB+E$;\par
    Normalize $D^{i+1}=D^{i+1}/|D^{i+1}|$;\par
    $i=i+1$;\par
}
Extract node mapping from $\hat{D}$ using a greedy method or by solving a LAP;
\end{algorithm}

\hypertarget{algorithm-4}{%
\paragraph{\texorpdfstring{\textsc{Umeyama} algorithm}{ algorithm}}\label{algorithm-4}}

\cite{Umeyama} is a spectral approach to find approximate solutions to the graph matching problem.
Assuming eigendecompositions of adjacency matrices \(A\) and \(B\) as \(A=U_A\Lambda_AU_A^\top\) and \(B=U_B\Lambda_BU_B^\top\), let \(|U_A|\) and \(|U_B|\) be matrices which takes absolute values of each element of \(U_A\) and \(U_B\).
Such modification to the eigenvector matrices guarantees the uniqueness of eigenvector selection.
The global mapping is obtained by minimizing the differences between matched rows of \(U_A\) and \(U_B\):
\begin{align*}
    \hat{P}=\mathop{\mathop{\mathrm{argmin}}}_{P\in\Pi}\lVert |U_A|-P|U_B|\rVert_F=\mathop{\mathop{\mathrm{argmax}}}_{P\in\Pi}\mathrm{trace}(|U_B||U_A|^\top P)
\end{align*}

The \textsc{Umeyama} algorithm can be generalized to matching directed graphs by eigendecomposing the Hermitian matrices \(E_A\) and \(E_B\) derived from the asymmetric adjacency matrices of the directed graphs.
The Hermitian matrix for the adjacency matrix \(A\) is defined as \(E_A=A_S+iA_N\), where \(A_S=(A+A^\top)/2\) is a symmetric matrix, \(A_N=(A-A^\top)/2\) is a skew-symmetric matrix and \(i\) is the imaginary unit.
Similarly, we can define the Hermitian matrix for \(B\).
Assume the eigendecompositions of \(E_A\) and \(E_B\) as follows:
\begin{align*}
    E_A=W_A\Gamma_AW_A^*, \quad  E_B=W_B\Gamma_BW_B^*
\end{align*}
and we aim at searching for:
\begin{align*}
    \hat{P}=\mathop{\mathop{\mathrm{argmax}}}_{P\in\Pi}\mathrm{trace}(|W_B||W_A|^\top P)
\end{align*}
Note that the \textsc{Umeyama} algorithm works on the condition that two graphs are isomorphic or nearly isomorphic.

\begin{algorithm}
\caption{Umeyama Algorithm}
\label{alg:Umeyama}
\SetKwInOut{Input}{Input}
\SetKwInOut{Output}{Output}
\Input{$A,B$}
\Output{permutation matrix $P$}
Compute the eigendecompositions of $A$ and $B$: $A=U_A\Lambda_AU_A^\top$, $B=U_B\Lambda_BU_B^\top$ ;\par
Solve the LAP: $P=\mathop{\mathop{\mathrm{argmax}}}_{P\in\Pi}\mathrm{trace}(|U_B||U_A|^\top P)$;\par
\end{algorithm}

\hypertarget{auxiliary-graph-matching-tools}{%
\subsection{Auxiliary graph matching tools}\label{auxiliary-graph-matching-tools}}

\hypertarget{centering-technique}{%
\subsubsection{Centering technique}\label{centering-technique}}

Instead of encoding the non-adjacencies by zeros in the adjacency matrices, the centering technique \citep{centering} assigns negative values to such edges.
The first approach is encoding non-adjacent node-pairs as \(-1\) with centered adjacency matrices \(\tilde{A}=2A-\textbf{J}\) and \(\tilde{B}=2B-\textbf{J}\), where \(\textbf{J}\) is a matrix of all ones.
An alternative approach relies on modeling assumptions where the pair of graphs are correlated but do not share a global structure.
We match \(\tilde{A} = A-\Lambda_A\) and \(\tilde{B} = B-\Lambda_B\), where \(\Lambda\) is an \(n\text{-by-}n\) matrix with \(ij\)-th entry denoting an estimated marginal probability of an edge.
In general, \(\Lambda\) is unknown but there are methods in the literature to estimate \(\Lambda\).

Matching centered graphs changes the rewards for matching edges, non-edges, and the penalties for mismatches.
Adapting the centering technique for the problem at hand can be used to find specific types of correspondences.
This can also be combined with constructing multilayer networks out of single layer networks to match according to multiple criteria \citep{Lin, GRAMPA}.
The centering technique can be applied to any of the implemented graph matching algorithm.
It is especially useful when padding graphs with differing numbers of vertices to distinguish isolated vertices from padded vertices.

\hypertarget{padding-graphs-of-different-orders}{%
\subsubsection{Padding graphs of different orders}\label{padding-graphs-of-different-orders}}

Until this section, we have been considering matching two graphs whose vertex sets are of the same cardinality.
However, matching graphs with different orders are commonly seen in real-world problems.

Suppose \(A\in\{0,1\}^{n\times n}\) and \(B\in\{0,1\}^{n_c\times n_c}\) with \(n_c<n\).
One can then pad the smaller graph with extra vertices to match the order of the larger graph, \(\tilde{B}=B\oplus \textbf{0}_{n-n_c}\) and match \(A\) and \(\tilde{B}\).
Every implemented graph matching algorithm in the \pkg{iGraphMatch} package automatically handles input graphs with a different number of vertices using sparse padding with minimal memory impact.

Since the isolated vertices and the padded vertices share the same topological structure, it can be useful to center the original graphs first then pad the smaller graph in the same manner.
This approach serves to differentiate between isolated vertices the padded ones.
It's theoretically verified that in the correlated {Erd{\H o}s-R{\'e}nyi} graph model, the centered padding scheme is guaranteed to find the true correspondence between the nodes of \(G_1\) and the induced subgraph of \(G_2\) under mild conditions even if \(|V_1|\ll|V_2|\), but the true alignment is not guaranteed without centering \citep{centering}.

\hypertarget{exploiting-sparse-and-low-rank-structure}{%
\subsubsection{Exploiting sparse and low-rank structure}\label{exploiting-sparse-and-low-rank-structure}}

Many real-world graphs, especially large graphs, are often very sparse with \(o(n^2)\) and often \(\theta(n)\) edges.
This can increase the difficulty of the graph matching problem due to the fact that there are fewer potential edges to match, but sparse graphs also come with computational advantages.
We rely on \pkg{igraph} and \pkg{Matrix} for efficient storage of these structures as well as the efficient implementation of various matrix operations.
We also use the LAPMOD algorithm for sparse LAP problems \citep{Volgenant1996-o} (see below).

Similarly, a low-rank structure appears in some of the procedures including starting at the rank-1 barycenter matrix and the different centering schemes.
Since low-rank matrices are generally not sparse and visa-versa we implemented the \texttt{splr} \proglang{S}4 class, standing for sparse plus low-rank matrices.
This class inherits from the \texttt{Matrix} class and includes slots for an \(n\times n\) sparse matrix \texttt{x} and \(n\times d\) dense matrices \texttt{a} and \texttt{b}, to represent matrices of the form \texttt{x\ +\ tcrossprod(a,\ b)}.
This class implements efficient methods for matrix multiplication and other operations that exploit the sparse and low-rank structure of the matrices.
Specifically, these methods often require only \(O(\|x\|_0) + O(nd)\) storage as opposed to \(O(n^2)\) required for densely stored matrices, and enjoy analogous computational advantages.
While users can also use these matrices explicitly, most use of them is automatic within functions such as \texttt{init\_start} and \texttt{center\_graph} and the matrices can largely be used interchangeably with other matrices.

\hypertarget{sssec:lap_methods}{%
\subsubsection{LAP methods}\label{sssec:lap_methods}}

Multiple graph matching methods include solving an LAP and so we have included multiple methods for solving LAPs into the package.
Specifically we implement the Jonker-Volgenant algorithm \citep{lap_solver} for dense cost matrices and the LAPMOD algorithm \citep{Volgenant1996-o} for sparse cost matrices.
Both algorithms are implemented in \proglang{C} to provide improved performance.
The LAPMOD approach is typically advantageous when the number of non-zero entries is less than 50\%.
We also depend on the \pkg{clue} package for the \texttt{solve\_LSAP} function which implements the Hungarian algorithm \citep{Papadimitriou1998-lz} for solving an LAP.
Each of these methods can be used independently of a specific graph matching method using the \texttt{do\_lap} function.

\hypertarget{multi-layered-graph-matching}{%
\subsubsection{Multi-layered graph matching}\label{multi-layered-graph-matching}}

Frequently, networks edges may have categorical attributes and from these categories, we can construct multilayer graphs \citep{multilayer}, where each layer in the networks contains edges from specific categories.
For matching two multilayer graphs, the standard graph matching problem can be extended as \(\sum_{l=1}^{m}\|A^{(l)} - PB^{(l)}P^\top\|_F^2\) where \(m\) denotes the number of categories and \(A^{(l)}, B^{(l)}\) are the adjacency matrices for the \(l\)th layers in each graph.
Note, we assume that the layers are aligned, so that layer \(l\) corresponds to the same edge-types in both multi-layer networks.

For an \pkg{igraph} object, the function \texttt{split\_igraph} can be used to convert a single object with categorical edge attributes into a list with each element only containing the edges with a specific attribute value.
The implemented algorithms can seamlessly match multi-layer graphs, which are encoded as a list of either \texttt{igraph} objects or \texttt{matrix}-like objects.
We also implemented a \texttt{matrixlist} \proglang{S}4 class that implements many standard matrix operations so that algorithms can be easily extended to work with multilayer networks.

\hypertarget{graph-models}{%
\subsection{Correlated random graph models}\label{graph-models}}

The correlated {Erd{\H o}s-R{\'e}nyi} model \citep{FAQ} is essential in the theoretical study of graph matching algorithms.
In a single {Erd{\H o}s-R{\'e}nyi} graph, each edge is present in the graph independently with probability \(p\).
The correlated {Erd{\H o}s-R{\'e}nyi} model provides a joint distribution for a pair of graphs, where each graph is marginally distributed as an {Erd{\H o}s-R{\'e}nyi} graph and corresponding edge-pairs are correlated.

\begin{defn}[Correlated {Erd{\H o}s-R{\'e}nyi} ]
Suppose $p \in (0,1)$ and $\rho\in [\max\{\frac{-p}{1-p}, \frac{p-1}{p}\},1]$, a pair of adjacency matrices $(A, B)\sim \text{CorrER}(pJ, \rho J)$ if:
For each $1\le u \le v \le n, A_{uv}$ are independent with $A_{uv} \sim Bernoulli(p)$.\;
For each $1\le u \le v \le n, B_{uv}$ are independent with $B_{uv} \sim Bernoulli(p)$.\;
$A_{uv}$ and $B_{u'v'}$ are independent unless $u=u', v=v'$ in which case the Pearson correlation $\mathrm{corr}(A_{uv},B_{uv})=\rho$.
\end{defn}

To sample a pair of correlated {Erd{\H o}s-R{\'e}nyi} graphs with edge probability \(p\), and Pearson correlation \(\rho\), we first sample three independent {Erd{\H o}s-R{\'e}nyi} graphs \(G_1\), \(Z_0\) and \(Z_1\) with edge probabilities \(p\), \(p(1-\rho)\) and \(p+\rho(1-p)\) respectively.
Let \(G_2 = (Z_1 \cap G_1) \bigcup (Z_0\cap G_1^c)\).

\cite{Percolation} provide an alternative formulation for the correlated {Erd{\H o}s-R{\'e}nyi} model.
First, one samples a single random {Erd{\H o}s-R{\'e}nyi} graph \(G\) with edge probability \(p'\).
Conditioned on \(G\), each edge in \(G\) is present independently in \(G_1,G_2\) with probability \(s'\).
These two parameterizations are related to each other by the relationship \(s'=p+\rho(1-p)\) and \(p'=p/(p+\rho(1-p))\).
The original parameterization is slightly more general because it allows for the possibility of negative correlation.

In addition to homogeneous correlated {Erd{\H o}s-R{\'e}nyi} random graphs, we also implement heterogeneous generalizations of this model.
The stochastic block model \citep{SBM} and the random dot product graphs (RDPG) model \citep{rdpg} can both be regarded as extensions of the {Erd{\H o}s-R{\'e}nyi} model.
The stochastic block model is useful to represent the community structure of graphs by dividing the graph into \(K\) groups.
Each node is assigned to a group and the probability of edges is determined by the group memberships of the vertex pair.
For the RDPG model, each vertex is assigned a latent position in \(\mathbb{R}^d\) and edge probabilities are given by the inner product between the latent positions of the vertex pair.

For both of these models, we can consider correlated graph-pairs where marginally they arise from one of these models and again corresponding edge pairs are correlated.

\hypertarget{sec:measure}{%
\subsection{Measures for goodness of matching}\label{sec:measure}}

The ability to assess the quality of the match when ground truth is unavailable is critical for the usage of the matching approaches.
There are various topological criteria that can be applied to measure the quality of matching results.
At the graph level, the most frequently used structural measures include matching pairs (MP), edge correctness (EC), and the size of the largest common connected subgraph (LCCS) \citep{measure}.
MP counts the number of correctly matched pairs of nodes, thus can only be used when the true alignment is available.
Global counts of common edges (CE) and common non-edges (CNE) can be defined as
\begin{align*}
    CE=\frac{1}{2}\sum_{i,j}\mathds{1}\{A_{ij}=PBP^\top_{ij} =1\}\quad CNE=\frac{1}{2}\sum_{i,j}\mathds{1}\{A_{ij}=PBP^\top_{ij} = 0\},
\end{align*}
along with error counts such extra edges (EE) and missing edges (ME),
\begin{align*}
    EE=\frac{1}{2}\sum_{i,j}\mathds{1}\{A_{ij}=0=1 - PBP^\top_{ij}\}\quad ME=\frac{1}{2}\sum_{i,j}\mathds{1}\{A_{ij}= 1 = 1 - PBP^\top_{ij}\},
\end{align*}

EC measures the percentage of correctly aligned edges, that is the fraction CE\(/|E_1|\).
The LCCS denotes the largest subset of aligned vertices such that the corresponding induced subgraphs of each graph are connected.
Matches with a larger LCCS are often preferable to those with many isolated components.

Another group of criteria measures the goodness of matching at the vertex level.
Informally, we aim at testing the hypotheses
\[H_0^{(v)}: \text{the vertex } v \text{ is not matched correctly by } P^*, \]
\[H_a^{(v)}: \text{the vertex } v \text{ is matched correctly by } P^*\]
for each vertex \(v\).

The goal is to address if the permutation matrix found by graph matching algorithm is significantly different from the one sampled from uniformly distributed permutation matrices \citep{row_perm}.
Unfortunately, vertex-level matching criteria have only received limited attention in the literature, however, we include two test statistics to measure fit.
The row difference statistic is the L\(_1\)-norm of the difference between \(A\) and \(P^*B{P^*}^\top\), namely
\begin{equation*}
T_d(v,P^*):=\Vert A_{v\cdot}-(P^*B{P^*}^\top)_{v\cdot}\Vert_1.
\end{equation*}
Intuitively, a correctly matched vertex \(v\) should induce a smaller \(T_d(v,P^*)\),
which for unweighted graphs corresponds to the number of edge disagreements induced by matching \(v\).
Alternatively, the row correlation statistic is defined as
\begin{equation*}
T_c(v,P^*):=1 - corr(A_{v\cdot},(P^*B{P^*}^\top)_{v\cdot}).
\end{equation*}
We expect the empirical correlation between the neighborhoods of \(v\) in \(A\) and \(P^*B{P^*}^\top\) to be larger for a correctly matched vertex.

We employ permutation testing ideas to the raw statistics as a normalization across vertices.
Let us take the row difference statistic for example.
The guiding intuition is that if \(v\) is correctly matched, the number of errors induced by \(P^*\) across the neighborhood of \(v\) in \(A\) and \(B\) (i.e., \(T_d(v, P^*)\)) should be significantly smaller than the number of errors induced by a randomly chosen permutation \(P\) (i.e., \(T_d(v, P)\)).

With this in mind, let \(\mathbb{E}_P\) and \(\mathrm{Var}_P\) denote the conditional expectation and variance of the raw statistic with \(P\) uniformly sampled over all permutation matrices.
The normalization is then given by
\begin{equation*}
T_p(v,P^*):=\frac{T(v,P^*)-\mathbb{E}_PT(v,P)}{\sqrt{Var_PT(v,P)}}
\end{equation*}
where \(T(v, P)\) can be either of the two test statistics we introduced earlier.

In addition to measuring match quality, these vertex-wise statistics can also serve as a tool to find which vertices have no valid match in another network, i.e., the vertex entity is present in one network but not the other.

\hypertarget{sec:usage}{%
\section{R functions and usage}\label{sec:usage}}

The \proglang{R} package \pkg{iGraphMatch} offers versatile options for graph matching and subsequent analysis.
Here we describe the usage of the package including sampling random correlated graph pairs, graph matching, and evaluating matching results.

\hypertarget{sampling-correlated-random-graph-pairs}{%
\subsection{Sampling correlated random graph pairs}\label{sampling-correlated-random-graph-pairs}}

We first illustrate the usage of functions for sampling correlated random graph pairs.
The usage of graph matching will be demonstrated on the graph-pairs sampled using these methods.

Functions of the form \texttt{sample\_correlated\_*\_pair} for sampling random graph pairs have the common syntax:

\begin{verbatim}
sample_correlated_*_pair(n, ***model parameters***,
  permutation = 1:n, directed = FALSE, loops = FALSE)
\end{verbatim}

The argument \texttt{n} specifies the number of nodes in each graph, and the default options are to sample a pair of undirected graphs without self-loop whose true alignment is the identity.
The \texttt{permutation} argument can be used to permute the vertex labels of the second graph.
The \texttt{model\ parameters} arguments vary according to different random graph models and typically consist of parameters for marginal graph distributions and for correlations between the corresponding edges.
The functions each return a named list of two \texttt{igraph} objects.

For the homogeneous correlated {Erd{\H o}s-R{\'e}nyi} graph model, the model parameters are \texttt{p}, the global edge probability, and \texttt{corr}, the Pearson correlation between aligned vertex-pairs.
For example, to sample a pair of graphs with 5 nodes from \(\mathrm{CorrER}(0.5J, 0.7J)\) we run

\begin{CodeChunk}
\begin{CodeInput}
R> library("iGraphMatch")
R> set.seed(1)
R> gnp_pair <- sample_correlated_gnp_pair(n = 5, corr = 0.7, p = 0.5)
R> (gnp_A <- gnp_pair$graph1)
\end{CodeInput}
\begin{CodeOutput}
IGRAPH ed00a0a U--- 5 4 -- Erdos renyi (gnp) graph
+ attr: name (g/c), type (g/c), loops (g/l), p (g/n)
+ edges from ed00a0a:
[1] 1--2 2--3 2--5 3--5
\end{CodeOutput}
\begin{CodeInput}
R> gnp_A[]
\end{CodeInput}
\begin{CodeOutput}
5 x 5 sparse Matrix of class "dgCMatrix"

[1,] . 1 . . .
[2,] 1 . 1 . 1
[3,] . 1 . . 1
[4,] . . . . .
[5,] . 1 1 . .
\end{CodeOutput}
\begin{CodeInput}
R> gnp_B <- gnp_pair$graph2
\end{CodeInput}
\end{CodeChunk}

Since we didn't obscure the vertex correspondence by assigning a value to the \texttt{permutation} argument, the underlying true alignment is the identity.

For the more general heterogeneous correlated {Erd{\H o}s-R{\'e}nyi} graph model, one needs to specify an edge probability matrix and a Pearson correlation matrix.
To sample a pair of graphs from the heterogeneous correlated {Erd{\H o}s-R{\'e}nyi} model again with 5 nodes in each graph, and with random edge probabilities and Pearson correlations:

\begin{CodeChunk}
\begin{CodeInput}
R> set.seed(123)
R> p <- matrix(runif(5 ^ 2, .5, .8), 5)
R> c <- matrix(runif(5 ^ 2, .5, .8), 5)
R> ieg_pair <- sample_correlated_ieg_pair(n = 5, p_mat = p, c_mat = c)
\end{CodeInput}
\end{CodeChunk}

Since the default is undirected graphs without self-loops, the entries of \texttt{p} and \texttt{c} along and below the diagonal are effectively ignored.

The stochastic block model requires block-to-block edge probabilities stored in the \texttt{pref.matrix} argument and the \texttt{block.sizes} argument indicates the size of each block, along with the Pearson correlation parameter \texttt{corr}.
Next, we sample a pair of graphs from the stochastic block model with two blocks of size 2 nodes and 3 nodes respectively, within-group edge probabilities of .7 and .5, across-group edge probability of .001, and Pearson correlation equal to .5.

\begin{CodeChunk}
\begin{CodeInput}
R> pm <- cbind(c(.7, .001), c(.001, .5))
R> sbm_pair <- sample_correlated_sbm_pair(n = 5, pref.matrix = pm,
+                                   block.sizes = c(2, 3), corr = 0.5)
\end{CodeInput}
\end{CodeChunk}

These functions also enables sampling a pair of correlated random graphs with junk vertices, i.e., vertices that don't have true correspondence in the other graph by specifying the number of overlapping vertices in the argument \texttt{ncore} or overlapping block sizes in the argument \texttt{core.block.sizes}.

The \pkg{iGraphMatch} package offers auxiliary tools for centering graphs to penalize the incorrect matches as well, which is implemented in the \texttt{center\_graph} function with syntax:

\begin{verbatim}
center_graph(A, scheme = c(-1, 1), use_splr = TRUE)
\end{verbatim}

with the first input being either a matrix-like or igraph object.
The \texttt{scheme} argument specifies the method for centering graphs.
Options include a pair of scalars where the entries of the adjacency matrix are linearly rescaled so that their minimum is \texttt{min(scheme)} and their maximum is \texttt{max(scheme)}.
Note, \texttt{scheme\ =\ "center"} is the same as \texttt{scheme\ =\ c(-1,\ 1)}.
Another option is to pass in a single integer, where the returned value is the adjacency matrix minus its best rank-\texttt{scheme} approximation.
The last argument \texttt{use\_splr} is a boolean indicating whether to return a \texttt{splrMatrix} object.
We use \texttt{use\_splr\ =\ FALSE} here to better display the matrices but \texttt{use\_splr\ =\ TRUE} will often result in improved performance, especially for large sparse networks.
Here, we center the sampled graph \texttt{gnp\_A} using different schemes:

\begin{CodeChunk}
\begin{CodeInput}
R> center_graph(gnp_A, scheme = "center", use_splr = FALSE)
\end{CodeInput}
\begin{CodeOutput}
5 x 5 Matrix of class "dgeMatrix"
     [,1] [,2] [,3] [,4] [,5]
[1,]   -1    1   -1   -1   -1
[2,]    1   -1    1   -1    1
[3,]   -1    1   -1   -1    1
[4,]   -1   -1   -1   -1   -1
[5,]   -1    1    1   -1   -1
\end{CodeOutput}
\begin{CodeInput}
R> center_graph(gnp_A, scheme = 2, use_splr = FALSE)
\end{CodeInput}
\begin{CodeOutput}
5 x 5 Matrix of class "dgeMatrix"
        [,1]    [,2]    [,3] [,4]    [,5]
[1,]  0.2068  0.0643 -0.0934    0 -0.0934
[2,]  0.0643  0.0200 -0.0290    0 -0.0290
[3,] -0.0934 -0.0290 -0.4578    0  0.5422
[4,]  0.0000  0.0000  0.0000    0  0.0000
[5,] -0.0934 -0.0290  0.5422    0 -0.4578
\end{CodeOutput}
\end{CodeChunk}

Users can then use the centered graphs as inputs to the implemented graph matching algorithms, which serve to alter rewards and penalties for common edges, common non-edges, missing edges, and extra edges.

\hypertarget{graph-matching-methods}{%
\subsection{Graph matching methods}\label{graph-matching-methods}}

The graph matching methods share the same basic syntax:

\begin{verbatim}
gm(A, B, seeds = NULL, similarity = NULL, method = "indefinite",
  ***algorithm parameters***)
\end{verbatim}

The first two arguments for graph matching algorithms represent two networks which can be matrices, igraph objects, or two lists of either form in the case of multi-layer matching.
The \texttt{seeds} argument contains prior information on the known partial correspondence of two graphs.
It can be a vector of logicals or indices if the seed pairs have the same indices in both graphs.
In general, the \texttt{seeds} argument takes a matrix or a data frame as input with two columns indicating the indices of seeds in the two graphs respectively.
The \texttt{similarity} parameter is for a matrix of similarity scores between the two vertex sets, with larger scores indicating higher similarity.
Notably, one should be careful with the different scales of the graph topological structure and the vertex similarity information in order to properly address the relative importance of each part of the information.

The \texttt{method} argument specifies a graph matching algorithm to use, and one can choose from ``indefinite'' (default), ``convex'', ``PATH'', ``percolation'', ``IsoRank'', ``Umeyama'', or a self-defined graph matching function which enables users to test out their own algorithms while remaining compatible with the package.
If \texttt{method} is a function, it should take at least two networks, seeds and similarity scores as arguments.
Users can also include additional arguments if applicable.
The self-defined graph matching function should return an object of the ``graphMatch'' class with matching correspondence, sizes of two input graphs, and other matching details.
As an illustrative example, \texttt{graph\_match\_rand} defines a new graph matching function which matches by randomly permuting the vertex label of the second graph using a random seed \texttt{rand\_seed}.
We then apply this self-defined GM method to matching the correlated {Erd{\H o}s-R{\'e}nyi} graphs sampled earlier with a specified random seed:

\begin{CodeChunk}
\begin{CodeInput}
R> graph_match_rand <- function(A, B, seeds = NULL,
+                              similarity = NULL, rand_seed){
+   totv1 <- nrow(A[[1]])
+   totv2 <- nrow(B[[1]])
+   nv <- max(totv1, totv2)
+
+   set.seed(rand_seed)
+   corr <- data.frame(corr_A = 1:nv,
+                      corr_B = c(1:nv)[sample(nv)])
+
+   graphMatch(
+     corr = corr,
+     nnodes = c(totv1, totv2),
+     detail = list(
+       rand_seed = rand_seed
+     )
+   )
+ }
R>
R> match_rand <- gm(gnp_A, gnp_B,
+                  method = graph_match_rand, rand_seed = 123)
\end{CodeInput}
\end{CodeChunk}

Other arguments vary for different graph matching algorithms with an overview given in Table\textasciitilde{}\ref{tab:gm-alg}.
The \texttt{start} argument for the \textsc{FW} methodology with ``indefinite'' and ``convex'' relaxations takes any \(nns\text{-by-}nns\) matrix or an initialization method including ``bari'', ``rds'' or ``convex''.
These represent initializing the iterations at a specific matrix, the barycenter, a random doubly stochastic matrix, or the doubly stochastic solution from ``convex'' method on the same graphs, respectively.

Moreover, sometimes we have access to side information on partial correspondence with uncertainty.
If we still treat such prior information as hard seeds and pass them through the \texttt{seeds} argument for ``indefinite'' and ``convex'' methods, incorrect information can yield unsatisfactory matching results.
Instead, we provide the option of soft seeding by incorporating the noisy partial correspondence into the initialization of the start matrix.
The core function used for initializing the start matrix with versatile options is the \texttt{init\_start} function.

\begin{longtable}[h]{llll}
\toprule\addlinespace
Parameter   & Type & Description   & Functions
\\\addlinespace
\midrule\endhead
\code{start}        & \begin{tabular}[c]{@{}l@{}} Matrix or\\ character\end{tabular} & \begin{tabular}[c]{@{}l@{}}Initialization of the start\\ matrix for iterations.\end{tabular}     & \code{FW, convex}
\\\addlinespace
\code{lap_method}   & Character   & Method for solving the LAP. & \begin{tabular}[c]{@{}l@{}}\code{FW, convex,}\\ \code{PATH, IsoRank}\end{tabular}
\\\addlinespace
\code{max_iter}     & Number      & \begin{tabular}[c]{@{}l@{}} Maximum number of\\ iterations.\end{tabular}  & \begin{tabular}[c]{@{}l@{}}\code{FW, convex,}\\ \code{PATH, IsoRank}\end{tabular}
\\\addlinespace
\code{tol} & Number       & \begin{tabular}[c]{@{}l@{}} Tolerance of edge\\ disagreements. \end{tabular} & \begin{tabular}[c]{@{}l@{}}\code{FW, convex,}\\ \code{PATH}\end{tabular}
\\\addlinespace
\code{r}  & Number      & \begin{tabular}[c]{@{}l@{}}Threshold of neighboring\\ pair scores. \end{tabular}    & \code{percolation}
\\\addlinespace
\begin{tabular}[c]{@{}l@{}}\code{ExpandWhen-}\\ \code{Stuck}\end{tabular}  & Boolean      & \begin{tabular}[c]{@{}l@{}l@{}} \code{TRUE} if performs\\ \code{ExpandWhenStuck}\\algorithm. \end{tabular} & \code{percolation}
\\\addlinespace
\bottomrule
\addlinespace
\caption{Overview of arguments for different graph matching functions.}
\label{tab:gm-alg}
\end{longtable}

Suppose the first two pairs of nodes are hard seeds and another pair of incorrect seed \((3,4)\) is soft seeds:

\begin{CodeChunk}
\begin{CodeInput}
R> hard_seeds <- 1:5 <= 2
R> soft_seeds <- data.frame(seed_A = 3, seed_B = 4)
\end{CodeInput}
\end{CodeChunk}

We generate a start matrix incorporating soft seeds initialized at the barycenter:

\begin{CodeChunk}
\begin{CodeInput}
R> as.matrix(start_bari <- init_start(start = "bari", nns = 3,
+       ns = 2, soft_seeds = soft_seeds))
\end{CodeInput}
\begin{CodeOutput}
     [,1] [,2] [,3]
[1,]  0.0    1  0.0
[2,]  0.5    0  0.5
[3,]  0.5    0  0.5
\end{CodeOutput}
\end{CodeChunk}

An alternative is to generate a start matrix that is a random doubly stochastic matrix incorporating soft seeds as follow

\begin{CodeChunk}
\begin{CodeInput}
R> set.seed(1)
R> as.matrix(start_rds <- init_start(start = "rds", nns = 3,
+       ns = 2, soft_seeds = soft_seeds))
\end{CodeInput}
\begin{CodeOutput}
      [,1] [,2]  [,3]
[1,] 0.000    1 0.000
[2,] 0.515    0 0.485
[3,] 0.485    0 0.515
\end{CodeOutput}
\end{CodeChunk}

Then we can initialize the \textsc{Frank-Wolfe} iterations at any of the start matrix by specifying the \texttt{start} parameter.

When there are no soft seeds, we no longer need to initialize the start matrix by using \texttt{init\_start} first. Instead we can directly assign an initialization method to the \texttt{start} argument in the \texttt{gm} function:

\begin{CodeChunk}
\begin{CodeInput}
R> match_rds <- gm(gnp_A, gnp_B, seeds = hard_seeds,
+                 method = "indefinite", start = "rds")
\end{CodeInput}
\end{CodeChunk}

Below use solution from the convex relaxation as the initialization for the indefinite relaxation.

\begin{CodeChunk}
\begin{CodeInput}
R> set.seed(123)
R> match_convex <- gm(gnp_A, gnp_B, seeds = hard_seeds,
+                    method = "indefinite", start = "convex")
\end{CodeInput}
\end{CodeChunk}

Now let's match the sampled pair of graphs \code{sbm_pair} from the stochastic block model by using \textsc{percolation} algorithm.
Apart from the common arguments for all the graph matching algorithms, \textsc{percolation} has another argument \code{r} representing the minimum number of matched neighbors required for matching a new qualified vertex pair.
Here we adopt the default value which is 2.
Also, at least one of similarity scores and seeds is required for \textsc{percolation} algorithm to kick off.
Let's utilize the same set of hard seeds and assume there is no available prior information on similarity scores.

\begin{CodeChunk}
\begin{CodeInput}
R> sbm_A <- sbm_pair$graph1
R> sbm_B <- sbm_pair$graph2
R> match_perco <- gm(sbm_A, sbm_A, seeds = hard_seeds,
+                   method = "percolation", r = 2)
R> match_perco
\end{CodeInput}
\begin{CodeOutput}
gm(A = sbm_A, B = sbm_A, seeds = hard_seeds, method = "percolation",
    r = 2)

Match (5 x 5):
  corr_A corr_B
1      1      1
2      2      2
\end{CodeOutput}
\end{CodeChunk}

Without enough prior information on partial correspondence, \textsc{percolation} couldn't find any qualifying matches.
Suppose in addition to the current pair of sampled graphs, the above sampled correlated homogeneous and heterogeneous {Erd{\H o}s-R{\'e}nyi} graphs are different layers of connectivity for the same set of vertices.
We can then match the nonseed vertices based on the topological information in all of these three graph layers.
To be consistent, let's still use the \textsc{percolation} algorithm with threshold \code{r} equal to 2 and the same set of seeds.

\begin{CodeChunk}
\begin{CodeInput}
R> lA <- list(sbm_A, ieg_pair$graph1, gnp_A)
R> lB <- list(sbm_B, ieg_pair$graph2, gnp_B)
R> match_perco_list <- gm(A = lA, B = lB, seeds = hard_seeds,
+                        method = "percolation", r = 2)
R> match_perco_list
\end{CodeInput}
\begin{CodeOutput}
gm(A = lA, B = lB, seeds = hard_seeds, method = "percolation",
    r = 2)

Match (5 x 5):
  corr_A corr_B
1      1      1
2      2      2
3      3      3
4      4      4
5      5      5
\end{CodeOutput}
\end{CodeChunk}

With the same amount of available prior information, we are now able to match all the nodes correctly.

Finally, we will give an example of matching multi-layers of graphs using \textsc{IsoRank} algorithm.
Unlike the other algorithm, similarity scores are required for \textsc{IsoRank} algorithm.
Without further information, we adopt the barycenter as the similarity matrix here.

\begin{CodeChunk}
\begin{CodeInput}
R> set.seed(1)
R> sim <- as.matrix(init_start(start = "bari", nns = 5,
+                             soft_seeds = hard_seeds))
R> match_IsoRank <- gm(A = lA, B = lB,
+                     seeds = hard_seeds, similarity = sim,
+                     method = "IsoRank", lap_method = "LAP")
\end{CodeInput}
\end{CodeChunk}

Graph matching functions return an object of class ``graphMatch'' which contains the details of the matching results, including a list of the matching correspondence, a call to the graph matching function and dimensions of the original two graphs.

\begin{CodeChunk}
\begin{CodeInput}
R> match_convex@corr
\end{CodeInput}
\begin{CodeOutput}
  corr_A corr_B
1      1      1
2      2      2
3      3      5
4      4      4
5      5      3
\end{CodeOutput}
\begin{CodeInput}
R> match_convex@call
\end{CodeInput}
\begin{CodeOutput}
gm(A = gnp_A, B = gnp_B, seeds = hard_seeds, method = "indefinite",
    start = "convex")
\end{CodeOutput}
\begin{CodeInput}
R> match_convex@nnodes
\end{CodeInput}
\begin{CodeOutput}
[1] 5 5
\end{CodeOutput}
\end{CodeChunk}

Additionally, ``graphMatch'' also returns a list of matching details corresponding to the specified method.
Table\textasciitilde{}\ref{tab:gm-value} provides an overview of returned values for different graph matching methods.
With the \texttt{seeds} information, one can obtain a node mapping for non-seeds accordingly

\begin{CodeChunk}
\begin{CodeInput}
R> match_convex[!match_convex$seeds]
\end{CodeInput}
\begin{CodeOutput}
  corr_A corr_B
3      3      5
4      4      4
5      5      3
\end{CodeOutput}
\end{CodeChunk}

The ``graphMatch'' class object can also be flexibly used as a matrix.
In addition to the returned list of matching correspondence, one can obtain the corresponding permutation matrix in the sparse form.

\begin{CodeChunk}
\begin{CodeInput}
R> match_convex[]
\end{CodeInput}
\begin{CodeOutput}
5 x 5 sparse Matrix of class "dgTMatrix"

[1,] 1 . . . .
[2,] . 1 . . .
[3,] . . . . 1
[4,] . . . 1 .
[5,] . . 1 . .
\end{CodeOutput}
\end{CodeChunk}

Notably, multiplicity is applicable to the ``graphMatch'' object directly without converting to the permutation matrix.
This enables obtaining the permuted second graph, that is \(PBP^\top\) simply by

\begin{CodeChunk}
\begin{CodeInput}
R> match_convex %*% gnp_B[]
\end{CodeInput}
\begin{CodeOutput}
5 x 5 sparse Matrix of class "dgCMatrix"

[1,] . 1 1 . .
[2,] 1 . 1 . 1
[3,] 1 1 . . 1
[4,] . . . . .
[5,] . 1 1 . .
\end{CodeOutput}
\end{CodeChunk}

\begin{longtable}[h]{lll}
\toprule\addlinespace
Parameter   & Description   & Functions
\\\addlinespace
\midrule\endhead
\code{seeds}        & \begin{tabular}[c]{@{}l@{}}A vector of logicals indicating if the\\ corresponding vertex is a seed.\end{tabular}     & All the functions.
\\\addlinespace
\code{soft}  & \begin{tabular}[c]{@{}l@{}l@{}}The functional similarity score matrix\\ with which one can extract more than\\ one matching candidates. \end{tabular}    & \begin{tabular}[c]{@{}l@{}}\code{FW, convex, PATH,}\\ \code{IsoRank, Umeyama}\end{tabular}
\\\addlinespace
\code{lap_method}      & Choice for solving the LAP. & \begin{tabular}[c]{@{}l@{}}\code{FW, convex,}\\ \code{IsoRank, Umeyama}\end{tabular}
\\\addlinespace
\code{iter}  & \begin{tabular}[c]{@{}l@{}}Number of iterations until convergence\\ or reaches the \code{max_iter}.\end{tabular} & \code{FW, convex, PATH}
\\\addlinespace
\code{max_iter}  & Maximum number of replacing matches. & \code{FW, convex}
\\\addlinespace
\code{match_order}  & The order of vertices getting matched. & \begin{tabular}[c]{@{}l@{}}\code{percolation,}\\ \code{IsoRank}\end{tabular}
\\\addlinespace
\bottomrule
\addlinespace
\caption{Overview of return values for different graph matching functions.}
\label{tab:gm-value}
\end{longtable}

\hypertarget{evaluation-of-goodness-of-matching}{%
\subsection{Evaluation of goodness of matching}\label{evaluation-of-goodness-of-matching}}

Along with the graph matching methodology, \pkg{iGraphMatch} has many capabilities for evaluating and visualizing the matching performance.
After matching two graphs, the function \texttt{summary} can be used to get a summary of the overall matching result in terms of commonly used measures including the number of matches, the number of correct matches, common edges, missing edges, extra edges and the objective function value.
The edge matching information is stored in a data frame named \texttt{edge\_match\_info}.
Note that \texttt{summary} outputs the number of correct matches only when the true correspondence is known by specifying the \texttt{true\_label} argument with a vector indicating the true correspondence in the second graph.
Applying the \texttt{summary} function on the matching result \texttt{match\_convex} with \texttt{true\_label\ =\ 1:5}, indicating the true correspondence is the identity that provides these summaries.

\begin{CodeChunk}
\begin{CodeInput}
R> summary(match_convex, gnp_A, gnp_B, true_label = 1:5)
\end{CodeInput}
\begin{CodeOutput}
Call: gm(A = gnp_A, B = gnp_B, seeds = hard_seeds, method = "indefinite",
    start = "convex")

# Matches: 3
# True Matches:  1, # Seeds:  2, # Vertices:  5, 5

  common_edges 4.00
 missing_edges 0.00
   extra_edges 1.00
         fnorm 1.41
\end{CodeOutput}
\end{CodeChunk}

Applying the \texttt{summary} function to a multi-layer graph matching result returns edge statistics for each layer.

\begin{CodeChunk}
\begin{CodeInput}
R> summary(match_perco_list, lA, lB)
\end{CodeInput}
\begin{CodeOutput}
Call: gm(A = lA, B = lB, seeds = hard_seeds, method = "percolation",
    r = 2)

# Matches: 3, # Seeds:  2, # Vertices:  5, 5
         layer    1    2    3
  common_edges 2.00 6.00 4.00
 missing_edges 0.00 1.00 0.00
   extra_edges 1.00 0.00 1.00
         fnorm 1.41 1.41 1.41
\end{CodeOutput}
\end{CodeChunk}

In realistic scenarios, the true correspondence is not available.
As introduced in Section \ref{sec:background}, the user can use vertex level statistics to evaluate match performance.
The \texttt{best\_matches} function evaluates a vertex-level metric and returns a sorted \texttt{data.frame} of the vertex-matches with the metrics.
The arguments are the two networks, a specific measure to use, the number of top-ranked vertex-matches to output, and the matching correspondence in the second graph if applicable.
As an example here, we apply \texttt{best\_matches} to rank the matches from above with the true underlying alignment

\begin{CodeChunk}
\begin{CodeInput}
R> best_matches(gnp_A, gnp_B, match = match_convex,
+              measure = "row_perm_stat", num = 3,
+              true_label = 1 : igraph::vcount(gnp_A))
\end{CodeInput}
\begin{CodeOutput}
  A_best B_best measure_value precision
1      4      4         -1.41     1.000
2      3      5         -1.22     0.500
3      5      3         -1.22     0.333
\end{CodeOutput}
\end{CodeChunk}

Note, \texttt{best\_matches} uses seed information from the \texttt{match} parameter and only outputs non-seed matches.
Without the true correspondence, \texttt{true\_label} would take the default value and the output data frame only contains the first three columns.

To visualize the matches of smaller graphs, the function \texttt{plot} displays edge discrepancies of the two matched graphs by an adjacency matrix or a ball-and-stick plot, depending on the input format of two graphs.

\begin{CodeChunk}
\begin{CodeInput}
R> plot(gnp_A, gnp_B, match_convex)
R> plot(gnp_A[], gnp_B[], match_convex)
\end{CodeInput}
\begin{figure}

{\centering \includegraphics[width=0.475\linewidth]{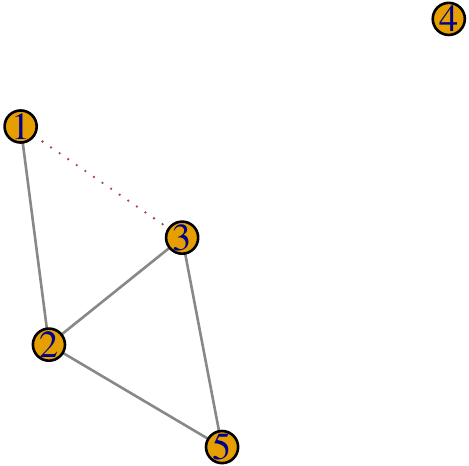} \includegraphics[width=0.475\linewidth]{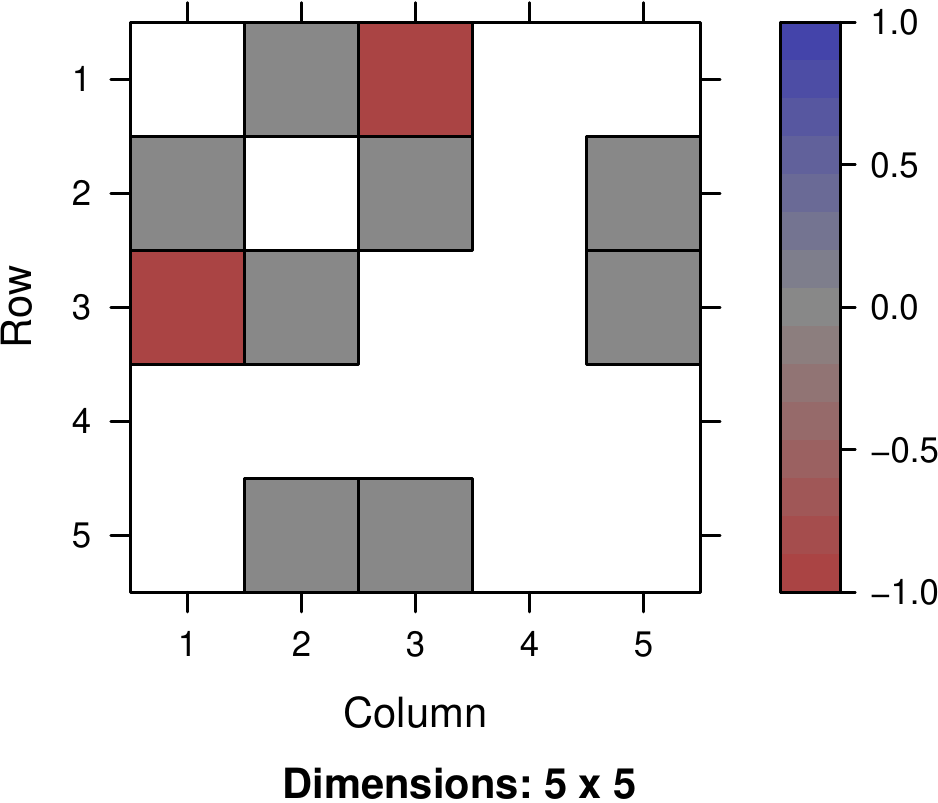}

}

\caption{\label{fig:visualization} Match visualizations. Grey, blue, and red colors indicate common edges, missing edges present only in the first network, and extra edges present only in the second network, respectively.}\label{fig:unnamed-chunk-2}
\end{figure}
\end{CodeChunk}

The plots for visualizing matching performance of \texttt{match\_convex} are shown in Figure\textasciitilde{}\ref{fig:visualization}.
Grey edges and pixels indicate common edges, red ones indicate edges only in the second graph.
If they were present, blue pixels and edges represent missing edges that only exist in the first graph.
The corresponding linetypes are solid, short dash, and long dash.

\hypertarget{sec:example}{%
\section{Examples}\label{sec:example}}

In this section, we demonstrate graph matching analysis using \pkg{iGraphMatch} via examples on real datasets, including communication networks, neuronal networks, and transportation networks.
Table \ref{tab:dataset-overview} presents brief overviews of the first two datasets.
Note that the number of edges doesn't consider weights for weighted graphs, and for directed graphs, an edge from node \(i\) to node \(j\) and another edge from \(j\) to \(i\) will be counted as two edges.
Tables \ref{tab:edge-summary} and \ref{tab:edge-summary-trans} summarize the edge correspondence between two graphs under the true alignment including the number of common edges, missing edges, and extra edges in two graphs.

In the first Enron email network example, we demonstrate the usage of \textsc{Frank-Wolfe} methodology and how to improve matching performance by using the centering technique and incorporating adaptive seeds.
In the second example using \emph{C. Elegans} synapses networks, we illustrate how to use soft matching for a challenging graph matching task using \textsc{Frank-Wolfe} methodology, \textsc{PATH} algorithm and \textsc{IsoRank} algorithm.
Finally, we include an example of matching two multi-layer graphs with similarity scores on the Britain transportation networks.

\begin{table}[ht]
\centering
\begin{tabular}{lllrlll}
  \hline
Dataset & Nodes & Edges & Correlation & Weighted & Directed & Loop \\
  \hline
Enron & 184 & 488 / 482 & 0.85 & No & Yes & Yes / Yes \\
  C. Elegans & 279 & 2194 / 1031 & 0.10 & Yes & Yes & No / Yes \\
   \hline
\end{tabular}
\caption{Overview of the Enron and C. Elegans graphs. \label{tab:dataset-overview}}
\end{table}

\begin{table}[ht]
\centering
\begin{tabular}{lrrr}
  \hline
Dataset & Common & Missing & Extra \\
  \hline
Enron & 412.00 & 76.00 & 70.00 \\
  C. Elegans & 116.00 & 981.00 & 399.50 \\
   \hline
\end{tabular}
\caption{Edge summary under the true alignments of the Enron and C. Elegans graphs. The columns indicate the number of common edges, missing edges in $G_1$, and extra edges in $G_2$. For weighted graphs, we define a pair of corresponding edges as a common edge as long as they both have positive weights. \label{tab:edge-summary}}
\end{table}

\hypertarget{sec:Enron}{%
\subsection{Example: Enron Email Network Data}\label{sec:Enron}}

The Enron email network data was originally made public by the Federal Energy Commission during the investigation into the Enron Corporation \citep{Enron}.
Each node of Enron network represents an email address and if there is at least one email sent from one address to another address, a directed edge exists between the corresponding nodes.

The \pkg{iGraphMatch} package includes the Enron email network data in the form of a pair of \texttt{igraph} objects derived from the original data where each graph represents one week of emails between 184 email addresses.
The two networks are unweighted and directed with edge densities around
0.014 in each graph and the empirical correlation between two graphs is 0.85.

First, let's load packages required for the following analysis:

\begin{CodeChunk}
\begin{CodeInput}
R> library("igraph")
R> library("iGraphMatch")
R> library("purrr")
R> library("dplyr")
\end{CodeInput}
\end{CodeChunk}

\hypertarget{visualization-of-enron-networks}{%
\subsubsection{Visualization of Enron networks}\label{visualization-of-enron-networks}}

We visualize the aligned Enron networks using the \code{plot} function with vertices sorted by a community detection algorithm \citep{community_detection} and degree.

For detailed interpretations to Figure\textasciitilde{}\ref{fig:Enron_graph}, please refer to Figure\textasciitilde{}\ref{fig:visualization}.

\begin{CodeChunk}
\begin{CodeInput}
R> g <- igraph::as.undirected(Enron[[1]])
R> com <- igraph::membership(igraph::cluster_fast_greedy(g))
R> deg <- rowSums(as.matrix(g[]))
R> ord <- order(max(deg) * com + deg)
R> plot(Enron[[1]][][ord, ord], Enron[[2]][][ord, ord])
\end{CodeInput}
\begin{figure}

{\centering \includegraphics[width=0.6\textwidth]{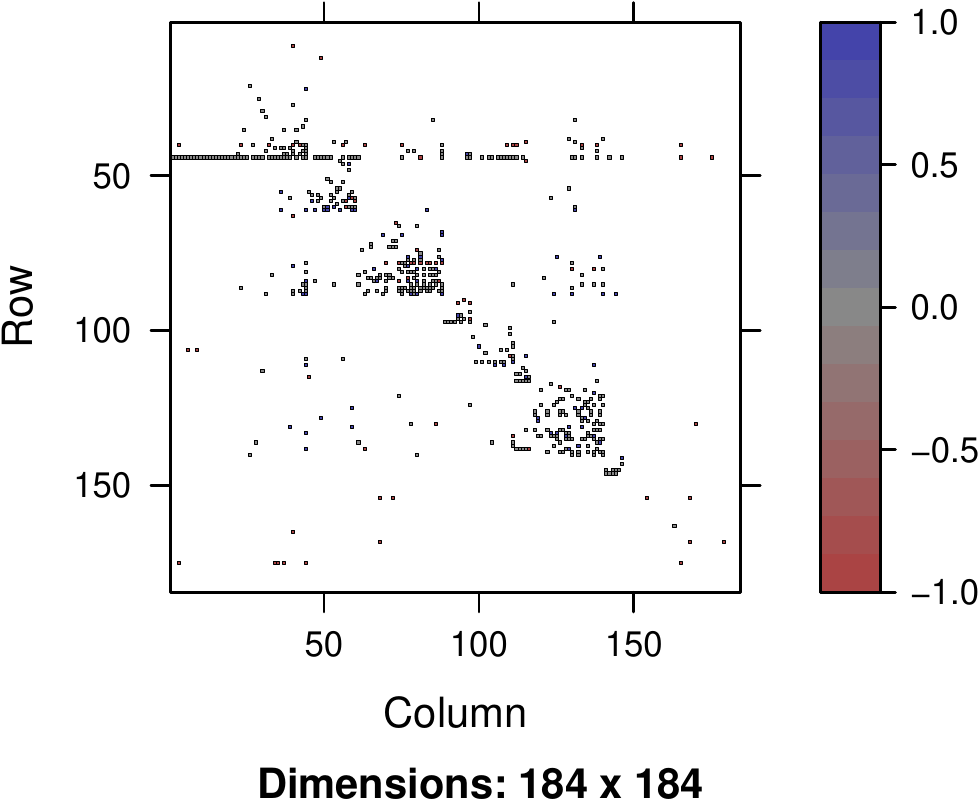}

}

\caption{\label{fig:Enron_graph} Asymmetric adjacency matrices of aligned Enron Corporation communication networks. The vertices are sorted by a community detection algorithm \citep{community_detection} and degree.}\label{fig:unnamed-chunk-3}
\end{figure}
\end{CodeChunk}

Note that 37 and 32 out of the total 184 nodes are isolated from the other nodes in two graphs respectively, indicating the corresponding employees haven't sent or received emails from other employees.
This adds difficulty to matching since it's impossible to distinguish the isolated nodes based on topological structure alone.
We first keep only the largest connected component of each graph.

\begin{CodeChunk}
\begin{CodeInput}
R> vid1 <- which(largest_cc(Enron[[1]])$keep)
R> vid2 <- which(largest_cc(Enron[[2]])$keep)
R> vinsct <- intersect(vid1, vid2)
R> v1 <- setdiff(vid1, vid2)
R> v2 <- setdiff(vid2, vid1)
R> A <- Enron[[1]][][c(vinsct, v1), c(vinsct, v1)]
R> B <- Enron[[2]][][c(vinsct, v2), c(vinsct, v2)]
\end{CodeInput}
\end{CodeChunk}

The sizes of largest connect components of two graphs are 146 and 151, which are different.
We reorder two graphs so that the first 145 nodes are aligned and common to both graphs.

\hypertarget{matching-largest-connected-components-using-fw-algorithm}{%
\subsubsection{Matching largest connected components using FW Algorithm}\label{matching-largest-connected-components-using-fw-algorithm}}

Let's assume the Enron email communication network from the second week is anonymous, and we aim at
finding an alignment between the email addresses from the first network and the second one to de-anonymize the latter.
Additionally, we want to find the email addresses that are active in both months.

Suppose no prior information on partial alignment is available in this example.
We match the two largest connected components using the \textsc{FW} algorithm with indefinite relaxation since seeds and similarity scores are not mandatory for this method.

Without any prior information, \code{seeds} and \code{similarity} arguments take default values which are \code{NULL}.
For the \code{start} argument, we assign equal likelihood to all the possible matches by initializing at the barycenter.
Since two graphs are of different sizes, the \code{gm} function automatically pads the smaller graph with extra 0's.

\begin{CodeChunk}
\begin{CodeInput}
R> set.seed(11)
R> match_FW <- gm(A = A, B = B, start = "bari", max_iter = 200)
R> head(match_FW)
\end{CodeInput}
\begin{CodeOutput}
  corr_A corr_B
1      1     95
2      2      2
3      3    140
4      4      4
5      5      5
6      6      6
\end{CodeOutput}
\end{CodeChunk}

Then, we check the summary of matching performance in terms of matched nodes, matched edges and the graph matching objective function.

\begin{CodeChunk}
\begin{CodeInput}
R> summary(match_FW, A, B)
\end{CodeInput}
\begin{CodeOutput}
Call: gm(A = A, B = B, start = "bari", max_iter = 200)

# Matches: 151, # Seeds:  0, # Vertices:  146, 151

  common_edges 379.0
 missing_edges 108.0
   extra_edges 102.0
         fnorm  14.5
\end{CodeOutput}
\end{CodeChunk}

In this example, we can evaluate the matching result based on statistics on matched edges.
Without any seeds or similarity scores, around 78\% of edges are correctly matched.

\hypertarget{centering-the-larger-graph}{%
\subsubsection{Centering the larger graph}\label{centering-the-larger-graph}}

We can try to improve performance by centering \texttt{B} by assigning -1 to non-edges, so that we penalize edges that are missing in \texttt{B} but present in \texttt{A}.

\begin{CodeChunk}
\begin{CodeInput}
R> A_center <- center_graph(A = A, scheme = "naive", use_splr = TRUE)
R> B_center <- center_graph(A = B, scheme = "center", use_splr = TRUE)
R> set.seed(11)
R> match_FW_center <- gm(A = A_center, B = B_center,
+                            start = "bari", max_iter = 200)
R> summary(match_FW_center, A, B)
\end{CodeInput}
\begin{CodeOutput}
Call: gm(A = A_center, B = B_center, start = "bari", max_iter = 200)

# Matches: 151, # Seeds:  0, # Vertices:  146, 151

  common_edges 393.0
 missing_edges  94.0
   extra_edges  88.0
         fnorm  13.5
\end{CodeOutput}
\end{CodeChunk}

From the summary tables, we would prefer matching Enron networks with the application of the centering scheme, since we get more matched common edges, as well as fewer missing edges and extra edges.

\hypertarget{matching-with-adaptive-seeds}{%
\subsubsection{Matching with adaptive seeds}\label{matching-with-adaptive-seeds}}

Supposing we have no access to ground truth, we use the \code{best_matches} function to measure and rank the vertex-wise matching performance.
Below shows the 6 matches that minimize the row permutation statistic.

\begin{CodeChunk}
\begin{CodeInput}
R> bm <- best_matches(A = A, B = B, match = match_FW_center,
+              measure = "row_perm_stat")
R> head(bm)
\end{CodeInput}
\begin{CodeOutput}
     A_best B_best measure_value
V83      65     65        -40.63
V75      57     57         -3.38
V147    115    115         -3.15
V59      43     43         -2.94
V64      48     48         -2.32
V51      36     36         -1.94
\end{CodeOutput}
\end{CodeChunk}

Since seeded graph matching enhances the graph matching performance substantially \citep{SGM}, it may be useful to use some of these best matches as seeds to improve matching results.
Here, we use adaptive seeds, taking the \(ns\) best matches and using them as seeds in a second run of the matching algorithm.
The table below displays edge statistics and objective function values for different number of adaptive seeds used.
The second column in the table shows the matching precision of the adaptive seeds based on ground truth.
Incorporating adaptive seeds and repeating the \textsc{FW} matching procedure on centered graphs further improve the matching results, compared with the case without any adaptive seeds when \(ns=0\).
The first 20 pairs of matched nodes ranked by \code{best_matches} function are all correctly matched, and this is also when matching is improved the most.

\begin{CodeChunk}
\begin{CodeInput}
R> match_w_hard_seeds <- function(ns){
+   seeds_bm <- head(bm, ns)
+   precision <- mean(seeds_bm$A_best == seeds_bm$B_best)
+   match_FW_center_seeds <- gm(A = A_center, B = B_center,
+                            seeds = seeds_bm[, 1:2], similarity = NULL,
+                            start = "bari", max_iter = 200)
+   edge_info <- summary(match_FW_center_seeds, A, B)$edge_match_info
+   cbind(ns, precision, edge_info)
+ }
R> set.seed(11)
R> map_dfr(seq(from = 0, to = 80, by = 20), match_w_hard_seeds)
\end{CodeInput}
\end{CodeChunk}

\begin{CodeChunk}

\begin{tabular}{rrrrrr}
\toprule
Seeds & Precision & Common & Missing & Extra & Fnorm\\
\midrule
0 & NaN & 393 & 94 & 88 & 13.5\\
20 & 1.00 & 402 & 85 & 79 & 12.8\\
40 & 1.00 & 410 & 77 & 71 & 12.2\\
60 & 0.95 & 401 & 86 & 80 & 12.9\\
80 & 0.92 & 399 & 88 & 82 & 13.0\\
\bottomrule
\end{tabular}

\end{CodeChunk}

As the number of adaptive seeds increases, the precision of adaptive seeds decreases.
Note that if they are treated as hard seeds, incorrect matches will remain in the matched set and might cause a cascade of errors.
An alternative way is to treat the top-ranked matches as soft seeds embedded in the start matrix to handle the uncertainty.
In this way, adaptive seeds not only provide prior information but also evolve over iterations.
The table below shows that the soft seeding approach always outperforms or performs as good as the hard seeding approach regardless of the number of adaptive seeds being used.

\begin{CodeChunk}
\begin{CodeInput}
R> match_w_soft_seeds <- function(ns){
+   seeds_bm <- head(bm, ns)
+   precision <- mean(seeds_bm$A_best == seeds_bm$B_best)
+   start_soft <- init_start(start = "bari",
+                            nns = max(dim(A)[1], dim(B)[1]),
+                            soft_seeds = seeds_bm[, 1:2])
+   match_FW_center_soft <- gm(A = A_center, B = B_center,
+                            start = start_soft, max_iter = 200)
+   edge_info <- summary(match_FW_center_soft, A, B)$edge_match_info
+   cbind(ns, precision, edge_info)
+ }
R> set.seed(11)
R> map_dfr(seq(from = 0, to = 80, by = 20), match_w_soft_seeds)
\end{CodeInput}
\end{CodeChunk}

\begin{CodeChunk}

\begin{tabular}{rrrrrr}
\toprule
Seeds & Precision & Common & Missing & Extra & Fnorm\\
\midrule
0 & NaN & 393 & 94 & 88 & 13.5\\
20 & 1.00 & 410 & 77 & 71 & 12.2\\
40 & 1.00 & 410 & 77 & 71 & 12.2\\
60 & 0.95 & 397 & 90 & 84 & 13.2\\
80 & 0.92 & 401 & 86 & 80 & 12.9\\
\bottomrule
\end{tabular}

\end{CodeChunk}

\hypertarget{core-vertices-detection}{%
\subsubsection{Core vertices detection}\label{core-vertices-detection}}

The function \code{best_matches} can also be used to detect core vertices.
Suppose the ground truth is known and that the first 145 vertices are core vertices.
The mean precision of detecting core vertices and junk vertices using \code{best_matches} function is displayed in Figure\textasciitilde{}\ref{fig:core}.
A lower rank is a stronger indicator of a core vertex and a higher rank is a stronger indicator of a junk vertex.
Let \(r^C_i, 1\le i\le n_c\) and \(r^J_j, 1\le j\le n_j\) denote the ranks associated with each core vertex and each junk vertex.
The figure shows the precision of identifying core vertices at each low rank \(r\), i.e., \(\frac{1}{r}\sum_{i = 1}^{n_c}\mathbbm{1}_{r^C_i\le r}\),
and the precision of identifying junk vertices at each high rank \(r\), i.e., \(\frac{1}{r}\sum_{j = 1}^{n_j}\mathbbm{1}_{r^J_j\ge n_c+n_j-r}\), which are separated by the vertical lines.

\begin{CodeChunk}
\begin{CodeInput}
R> nc <- length(vinsct)
R> nj <- max(length(v1), length(v2))
R> core_precision <- map_dbl(1 : nc,
+                           ~mean(bm$A_best[1 : .x] <= nc))
R> junk_precision <- map_dbl(1 : nj,
+                           ~mean(bm$A_best[(nc + .x) : (nc + nj)] > nc))
\end{CodeInput}
\end{CodeChunk}

Core detection performance is substantially better than chance, as represented by the dotted horizontal lines.
The top 83 are all core vertices indicating good overall performance for core identification.
For junk identification, the junk vertices are ranked 68, 58, 54, 22, 15, 14 according to which have the lowest score, indicating that some junk vertices are difficult to identify.

\begin{CodeChunk}
\begin{figure}

{\centering \includegraphics[width=0.5\textwidth,height=0.24\textheight]{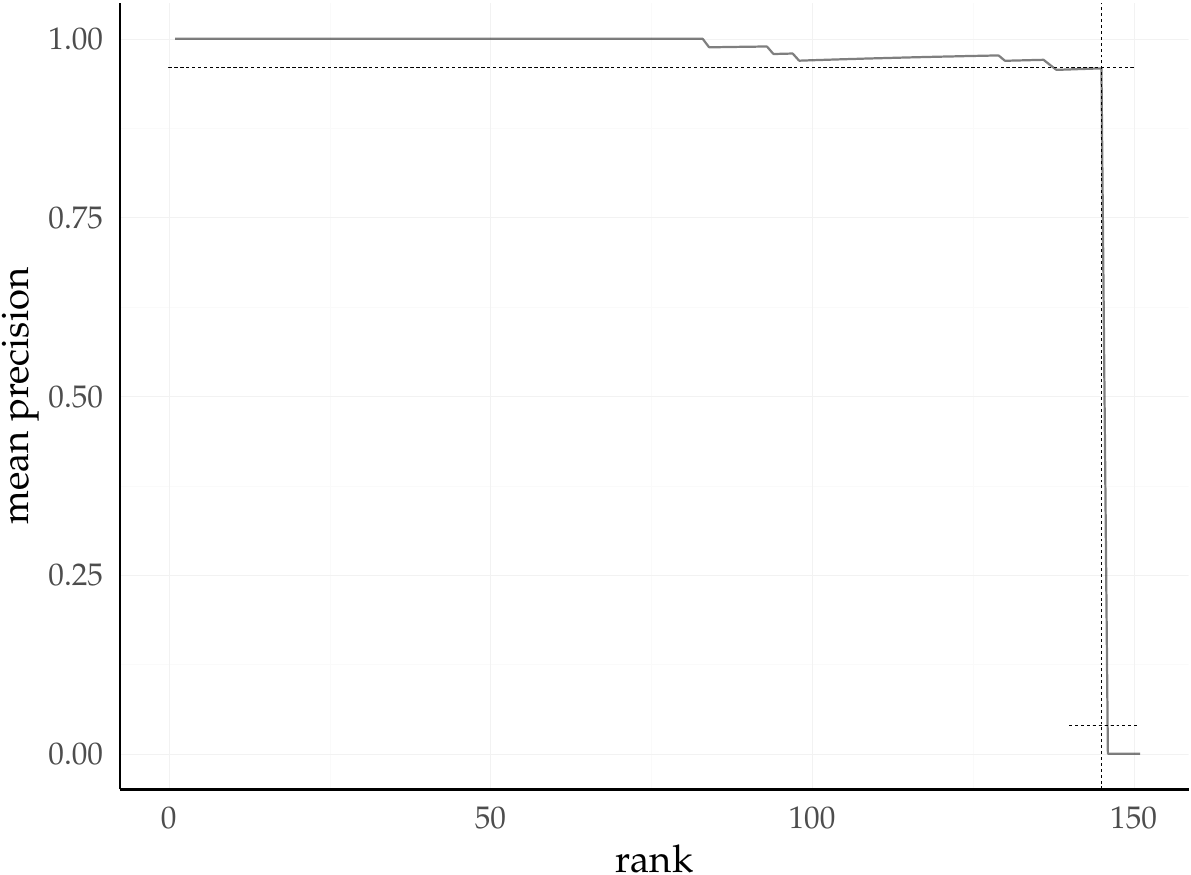}

}

\caption{\label{fig:core}Mean precision for identifying core and junk vertices for the Enron networks by using the row permutation test. The vertical lines separate the performance of identifying core vertices with low ranks from junk vertices with high ranks. The horizontal lines indicate the performance of a random classifier.}\label{fig:unnamed-chunk-4}
\end{figure}
\end{CodeChunk}

\hypertarget{sec:CE}{%
\subsection{Example: C. Elegans Network Data}\label{sec:CE}}

The \emph{C. Elegans} networks consist of the chemical synapses network and the electrical synapses network of the roundworm, where each of 279 nodes represents a neuron and each edge represents the intensity of synapse connections between two neurons \citep{neuro}.
Matching the chemical synapses network to the electrical synapses network is essential for understanding how the brain functions.
These networks are quite sparse with edge densities of 0.028 and 0.013 in each graph and the empirical correlation between two graphs is 0.1.

\hypertarget{a-challenging-task}{%
\subsubsection{A challenging task}\label{a-challenging-task}}

For simplicity, we made the networks unweighted and undirected for the experiments, and we assume the ground truth is known to be the identity.

\begin{CodeChunk}
\begin{CodeInput}
R> C1 <- C.Elegans[[1]][] > 0
R> C2 <- C.Elegans[[2]][] > 0
R> plot(C1[], C2[])
R> match <- gm(C1, C2, start = Matrix::Diagonal(nrow(C1)))
R> plot(C1[], C2[], match)
\end{CodeInput}
\begin{figure}

{\centering \subfloat[True correspondence\label{fig:unnamed-chunk-5-1}]{\includegraphics[height=0.24\textheight,keepaspectratio]{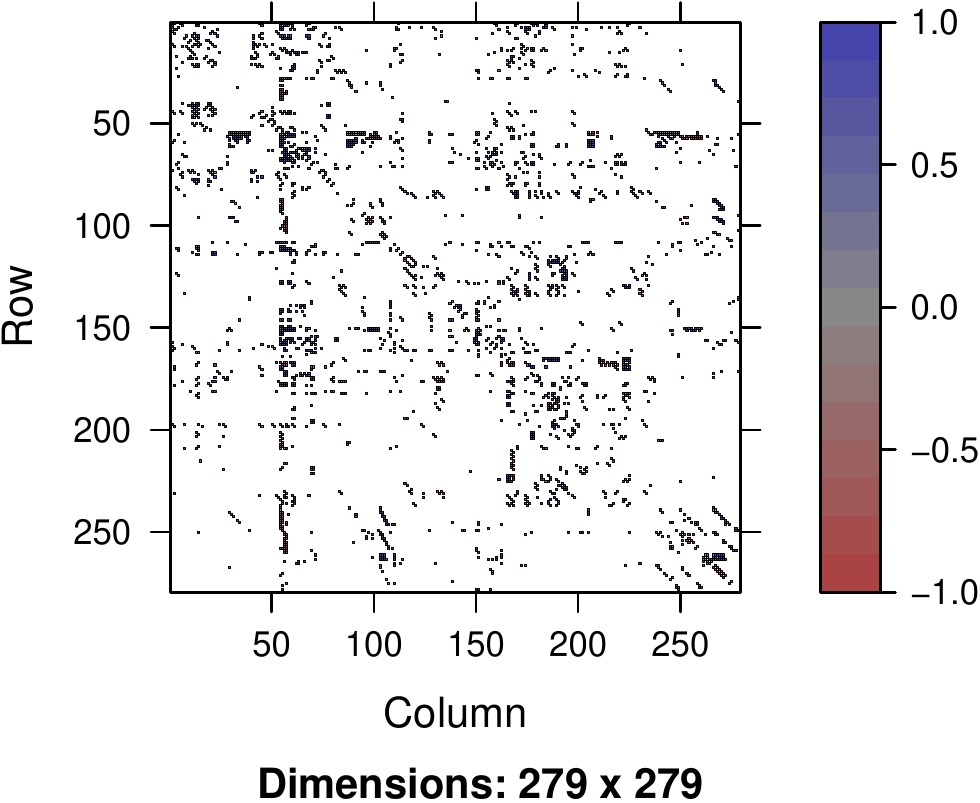} }\subfloat[Alignment found by FW initialized at true alignment\label{fig:unnamed-chunk-5-2}]{\includegraphics[height=0.24\textheight,keepaspectratio]{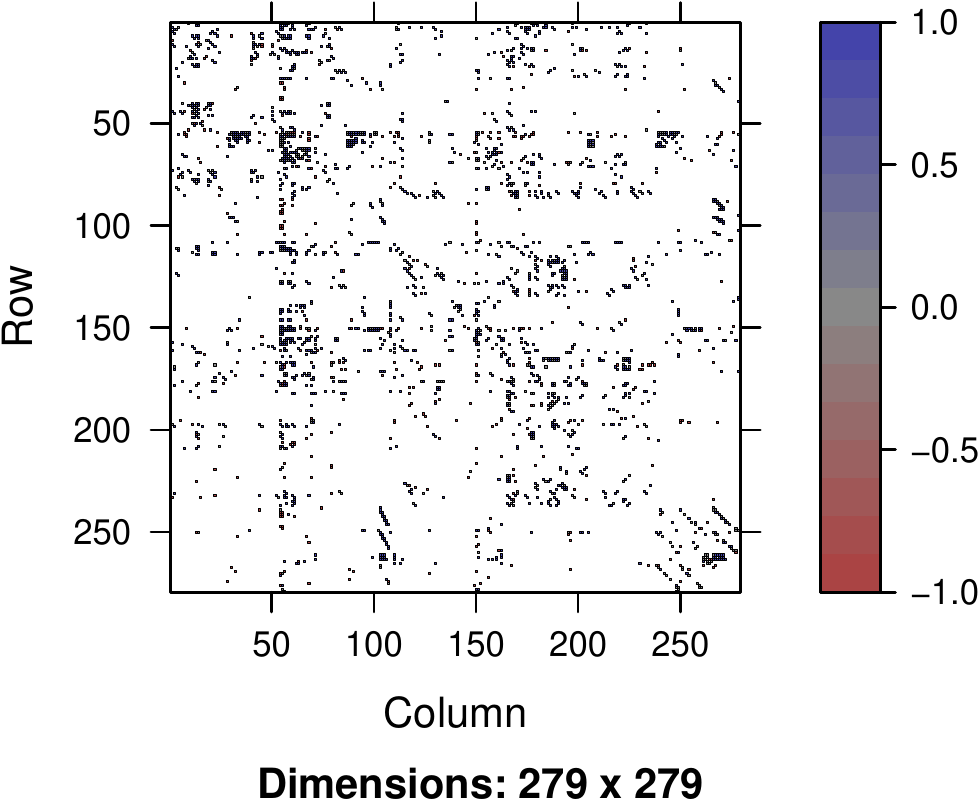} }

}

\caption{\label{fig:edge} Edge discrepancies for the matched graphs with the true correspondence and FW algorithm starting at the true correspondence. Green pixels represents an edge in the chemical graph while no edge in the electrical graph. Red pixels represent only an edge in the electrical graph. Grey pixels represent there is an edge in both graphs and white represents no edge in both graphs.}\label{fig:unnamed-chunk-5}
\end{figure}
\end{CodeChunk}

Matching the \emph{C. Elegans} networks is a challenging task.
Figures\textasciitilde{}\ref{fig:edge} depict the edge discrepancies of two networks under the true alignment and the matching correspondence using \textsc{FW} algorithm initialized at the true alignment.
The alignment found using \textsc{FW} is not the identity with 112 out of 279 nodes correctly matched
and improves upon the identity in terms of the number of edge discrepancies.
For the true alignment, there are
116
edge errors and
1380
common edges while the alignment yielded by \textsc{FW} initialized at the true correspondence has
269.5
edge errors and
1074
common edges.
Hence, this graph matching object does not have a solution at the true alignment.
One can try to use other objective functions to enhance the matching result, however we do not investigate this here.
Overall, while most performance measures are poor, our results illustrate the spectrum of challenges for graph matching.

\hypertarget{soft-matching-map3}{%
\subsubsection{Soft matching: MAP@3}\label{soft-matching-map3}}

Considering matching \emph{C. Elegans} graphs is quite challenging, let's assume 20 pairs of vertices are known as seeds, which are chosen at random.
Accordingly, we generate a similarity matrix with 1's corresponding to seeds, and the rest being barycenter.

\begin{CodeChunk}
\begin{CodeInput}
R> seeds <- sample(nrow(C1), 20)
R> sim <- init_start(start = "bari", nns = nrow(C1), soft_seeds = seeds)
\end{CodeInput}
\end{CodeChunk}

In addition to one-on-one matching, we will also conduct soft matching, which is to find three most promising matches to each non-seed vertex.
We achieve the goal of soft matching by finding the top 3 largest values in each row of the doubly stochastic matrix from the last iteration of \textsc{Frank Wolfe} methodology with indefinite relaxation and \textsc{PATH} algorithm, as well as the normalized matrix from the last iteration of the power method for \textsc{IsoRank} algorithm.
To evaluate the matching performance, we will look at both matching precision: \(precision=\frac{1}{n_m-s}\sum_{i\in V_m\setminus S}P_{ii}\), and Mean Average Precision @ 3 (MAP@ 3):\(MAP@3 = \frac{1}{n_m-s}\sum_{i\in V_m\setminus S}\mathbbm{1}_{\{i\in T_i\}}\), where \(T_i\) is the set of 3 most promising matches to node \(i\).

\begin{CodeChunk}
\begin{CodeInput}
R> set.seed(123)
R> m_FW <- gm(A = C1, B = C2, seeds = seeds,
+            similarity = sim, method = "indefinite",
+            start = "bari", max_iter = 100)
R> m_PATH <- gm(A = C1, B = C2, seeds = seeds,
+              similarity = NULL, method = "PATH",
+              epsilon = 1, tol = 1e-05)
R> m_Iso <- gm(A = C1, B = C2, seeds = seeds,
+             similarity = as.matrix(sim), method = "IsoRank",
+             max_iter = 50, lap_method = "LAP")
\end{CodeInput}
\end{CodeChunk}

\begin{CodeChunk}

\begin{tabular}{llll}
\toprule
  & Frank wolfe & PATH & IsoRank\\
\midrule
Precision & 0.1039 & 0.0824 & 0.0896\\
MAP3 & 0.1111 & 0.086 & 0.0932\\
\bottomrule
\end{tabular}

\end{CodeChunk}

MAP@ 3 is slightly higher than precision for each method.
Soft matching provides an alternative way of matching by generating a set of promising matching candidates.

\hypertarget{sec:Transp}{%
\subsection{Example: Britain Transportation Network}\label{sec:Transp}}

To demonstrate matching multi-layer networks-layers, we consider two graphs derived from the Britain Transportation network \citep{transportation}.
The network reflects the transportation connections in the UK, with five layers representing ferry, rail, metro, coach, and bus.
A smaller template graph was constructed based on a random walk starting from a randomly chosen hub node, a node that has connections in all the layers.
The template graph has 53 nodes and 56 connections in total and is an induced subgraph of the original graph.

Additionally, based on filter methods from \citet{Moorman2018-ha}, the authors of that paper also provided a list of candidate matches for each template node, where the true correspondence is guaranteed to be among the candidates.
The number of candidates ranges from 3 to 1059 at most, with an average of 241 candidates for each template vertex.
Thus, we made an induced subgraph from the transportation network with only candidates, which gave us the world graph with 2075 vertices and 8368 connections.

\begin{CodeChunk}
\begin{CodeInput}
R> tm <- Transportation[[1]]
R> cm <- Transportation[[2]]
R> candidate <- Transportation[[3]]
\end{CodeInput}
\end{CodeChunk}

Figure\textasciitilde{}\ref{Fig:trans_net} visualizes the transportation connections for the induced subgraphs, where means of transportation are represented by different colors.
Note that all edges in the template are common edges shared by two graphs,
where 40\%, 24.1\%, 37.5\%, 31.7\% and 25.6\% of edges in the world graph are in template for each layer.
All graphs are unweighted, directed, and do not have self-loops.
Tables \ref{tab:edge-summary-trans} further displays an overview and edge summary regarding each layer of the Britain Transportation Network.
A true correspondence exists for each template vertex in the world graph, our goal is to locate each template vertex in the Britain Transportation network by matching two multi-layer graphs with different number of vertices.

\begin{table}[ht]
\centering
\begin{tabular}{lllrrrr}
  \hline
Layer & Nodes & Edges & Correlation & Common & Missing & Extra \\
  \hline
Ferry & 53 / 2075 & 10 / 42 & 0.63 &  10 &   0 &  15 \\
  Rail & 53 / 2075 & 14 / 4185 & 0.49 &  14 &   0 &  44 \\
  Metro & 53 / 2075 & 9 / 445 & 0.61 &   9 &   0 &  15 \\
  Coach & 53 / 2075 & 13 / 2818 & 0.56 &  13 &   0 &  28 \\
  Bus & 53 / 2075 & 10 / 878 & 0.50 &  10 &   0 &  29 \\
   \hline
\end{tabular}
\caption{Overview of the Britain Transportation Network layers. Correlation is calculted using the template graph and the aligned induced subgraph of the world graph. The final three columns indicate the number of common edges, missing edges, and extra edges in the aligned subgraph of the world graph. \label{tab:edge-summary-trans}}
\end{table}

\begin{CodeChunk}
\begin{figure}

{\centering \includegraphics[width=0.8\linewidth]{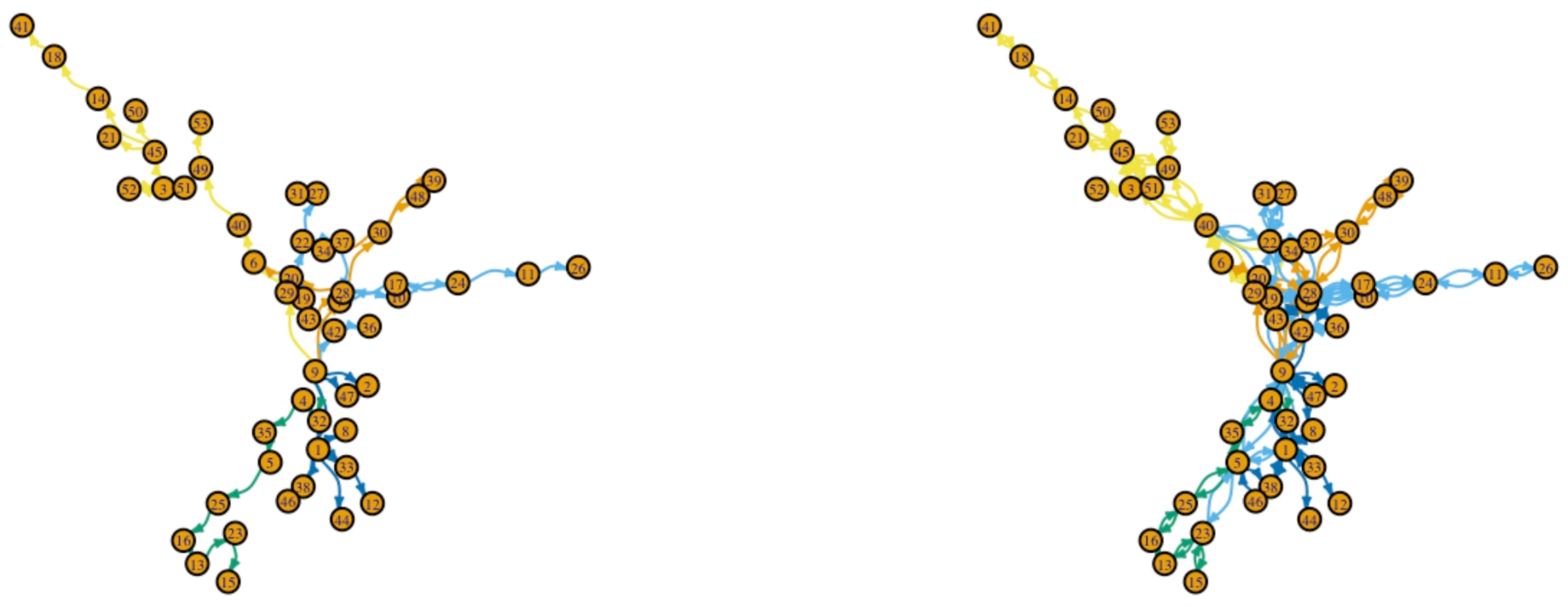}

}

\caption{\label{Fig:trans_net} Visualization of the template graph (left) and the world graph (right) with corresponding vertices, both derived from the Britain Transportation network with five layers: ferry, rail, metro, coach, and bus. Edges represent transportation transactions and each color indicates a different means of transportation from a different layer of network.}\label{fig:unnamed-chunk-6}
\end{figure}
\end{CodeChunk}

Based on the candidates, we specify a start matrix that is row-stochastic which can be used for the \code{start} argument in the graph matching function for \textsc{FW} methodology.
For each row node, its value is either zero or the inverse of the number of candidates for that node.
To ensure that template nodes only get matched to candidates, we constructed a similarity score matrix by taking the start matrix \(\times 10^5\), so that a high similarity score is assigned to all the template-candidate pairs.

Then we match the template graph with the world graph using \textsc{Percolation} algorithm.
The template graph stored in \code{tm} and world graph \code{cm} are lists of 5 matrices of dimensions 53 and 2075 respectively.
Since we have no information on seeds, we assign \code{NULL} to the \code{seeds} argument, the \textsc{Percolation} algorithm will initialize the mark matrix using prior information in the similarity score matrix.

\begin{CodeChunk}
\begin{CodeInput}
R> match <- gm(A = tm, B = cm, similarity = similarity,
+             method = "percolation", r = 4)
R> summary(match, tm, cm)
\end{CodeInput}
\begin{CodeOutput}
Call: gm(A = tm, B = cm, similarity = similarity, method = "percolation",
    r = 4)

# Matches: 53, # Seeds:  0, # Vertices:  53, 2075
         layer     1  2     3    4     5
  common_edges 10.00 13  9.00 12.0 10.00
 missing_edges  0.00  1  0.00  1.0  0.00
   extra_edges 22.00 35 21.00 25.0 35.00
         fnorm  4.69  6  4.58  5.1  5.92
\end{CodeOutput}
\end{CodeChunk}

The \code{summary} function outputs edge statistics and objective function values for each layer separately.
To further improve matching performance, one can replicate all the analysis in the first example on Enron dataset, such as using the centering scheme and adaptive seeds.
Finally, one can refer to the match report to compare matching performance and pick the best one.

\hypertarget{sec:conclusion}{%
\section{Conclusions}\label{sec:conclusion}}

In this work, we detail the methods and usage of the \proglang{R} package \pkg{iGraphMatch} for finding and assessing an alignment between the vertex sets of two edge-correlated graphs.
The package implements common steps for the analysis of graph matching: seamless matching of generalized graphs, evaluation of matching performance, and visualization.
For each of the graph matching methodologies, we provide versatile options for the form of input graphs and the specification of available prior information.
Through the discussion in Section \ref{sec:example}, we demonstrate the broad functionality and flexibility of the package by analyzing diverse graph matching problems on real data step by step.
The package also provides tools for simulating correlated graphs which can be used in the development and enhancement of graph matching methods.

Methods for graph matching are still under active development.
We plan to include other novel methods as the field continues to develop.
In the short term we are looking to introduce a suite of additional matching methods that have recently been proposed in the literature.

One of the biggest challenges for graph matching is evaluating the quality of a match, especially at the vertex level.
This has received minimal attention in the previous literature.
We provide measures of goodness of matching on the vertex level and demonstrate their effectiveness empirically.
These baseline methods implement a permutation testing framework for assessing matches that can be readily extended to other metrics.

\section*{Acknowledgments}

The primary authors for the package are Vince Lyzinski, Zihuan Qiao, and Daniel Sussman.
Joshua Agterberg, Lujia Wang, and Yixin Kong also provided important contributions.
We also want to thank all of our users, especially Youngser Park, for their feedback and patience as we continue to develop the package.

This work was supported in part by grants from DARPA (FA8750-20-2-1001 and FA8750-18-0035) and from MIT Lincoln Labs.

\bibliography{refs.bib}

\begin{thebibliography}{44}
\newcommand{\enquote}[1]{``#1''}
\providecommand{\natexlab}[1]{#1}
\providecommand{\url}[1]{\texttt{#1}}
\providecommand{\urlprefix}{URL }
\expandafter\ifx\csname urlstyle\endcsname\relax
  \providecommand{\doi}[1]{doi:\discretionary{}{}{}#1}\else
  \providecommand{\doi}{doi:\discretionary{}{}{}\begingroup
  \urlstyle{rm}\Url}\fi
\providecommand{\eprint}[2][]{\url{#2}}

\bibitem[{Aflalo \emph{et~al.}(2015{\natexlab{a}})Aflalo, Bronstein, and
  Kimmel}]{friendly}
Aflalo Y, Bronstein A, Kimmel R (2015{\natexlab{a}}).
\newblock \enquote{On convex relaxation of graph isomorphism.}
\newblock \emph{Proceedings of the National Academy of Sciences},
  \textbf{112}(10), 2942--2947.
\newblock ISSN 0027-8424.
\newblock \doi{10.1073/pnas.1401651112}.
\newblock \eprint{https://www.pnas.org/content/112/10/2942.full.pdf},
  \urlprefix\url{https://www.pnas.org/content/112/10/2942}.

\bibitem[{Aflalo \emph{et~al.}(2015{\natexlab{b}})Aflalo, Bronstein, and
  Kimmel}]{relax_paper}
Aflalo Y, Bronstein A, Kimmel R (2015{\natexlab{b}}).
\newblock \enquote{On convex relaxation of graph isomorphism.}
\newblock \emph{Proceedings of the National Academy of Sciences},
  \textbf{112}(10), 2942--2947.
\newblock ISSN 0027-8424.
\newblock \doi{10.1073/pnas.1401651112}.
\newblock \eprint{https://www.pnas.org/content/112/10/2942.full.pdf},
  \urlprefix\url{https://www.pnas.org/content/112/10/2942}.

\bibitem[{{Belongie} \emph{et~al.}(2002){Belongie}, {Malik}, and
  {Puzicha}}]{SimilarityScore}
{Belongie} S, {Malik} J, {Puzicha} J (2002).
\newblock \enquote{Shape matching and object recognition using shape contexts.}
\newblock \emph{IEEE Transactions on Pattern Analysis and Machine
  Intelligence}, \textbf{24}(4), 509--522.
\newblock \doi{10.1109/34.993558}.

\bibitem[{{Berg} \emph{et~al.}(2005){Berg}, {Berg}, and {Malik}}]{PatternRec1}
{Berg} AC, {Berg} TL, {Malik} J (2005).
\newblock \enquote{Shape matching and object recognition using low distortion
  correspondences.}
\newblock \textbf{1}, 26--33 vol. 1.
\newblock ISSN 1063-6919.
\newblock \doi{10.1109/CVPR.2005.320}.

\bibitem[{Burkard \emph{et~al.}(2009)Burkard, Dell'Amico, and Martello}]{AP}
Burkard R, Dell'Amico M, Martello S (2009).
\newblock \emph{Assignment Problems}.
\newblock Society for Industrial and Applied Mathematics, Philadelphia, PA,
  USA.
\newblock ISBN 0898716632, 9780898716634.

\bibitem[{{Caelli} and {Kosinov}(2004)}]{PatternRec2}
{Caelli} T, {Kosinov} S (2004).
\newblock \enquote{An eigenspace projection clustering method for inexact graph
  matching.}
\newblock \emph{IEEE Transactions on Pattern Analysis and Machine
  Intelligence}, \textbf{26}(4), 515--519.
\newblock ISSN 0162-8828.
\newblock \doi{10.1109/TPAMI.2004.1265866}.

\bibitem[{Chen \emph{et~al.}(2015)Chen, Vogelstein, Lyzinski, and
  Priebe}]{neuro}
Chen L, Vogelstein JT, Lyzinski V, Priebe CE (2015).
\newblock \enquote{A Joint Graph Inference Case Study: the C.elegans Chemical
  and Electrical Connectomes.}
\newblock \eprint{1507.08376}.

\bibitem[{Clauset \emph{et~al.}(2004)Clauset, Newman, and
  Moore}]{community_detection}
Clauset A, Newman MEJ, Moore C (2004).
\newblock \enquote{Finding community structure in very large networks.}
\newblock \emph{Physical Review E}, \textbf{70}(6).
\newblock ISSN 1550-2376.
\newblock \doi{10.1103/physreve.70.066111}.
\newblock \urlprefix\url{http://dx.doi.org/10.1103/PhysRevE.70.066111}.

\bibitem[{Conte \emph{et~al.}(2004)Conte, Foggia, Vento, and
  Sansone}]{PatternRec3}
Conte D, Foggia P, Vento M, Sansone C (2004).
\newblock \enquote{{Thirty Years Of Graph Matching In Pattern Recognition}.}
\newblock \emph{{International Journal of Pattern Recognition and Artificial
  Intelligence}}, \textbf{18}(3), 265--298.
\newblock \doi{10.1142/S0218001404003228}.
\newblock \urlprefix\url{https://hal.archives-ouvertes.fr/hal-01408706}.

\bibitem[{Cour \emph{et~al.}(2007)Cour, Srinivasan, and Shi}]{ML2}
Cour T, Srinivasan P, Shi J (2007).
\newblock \enquote{Balanced Graph Matching.}
\newblock pp. 313--320.
\newblock
  \urlprefix\url{http://papers.nips.cc/paper/2960-balanced-graph-matching.pdf}.

\bibitem[{Csardi and Nepusz(2006)}]{igraph}
Csardi G, Nepusz T (2006).
\newblock \enquote{The \pkg{igraph} software package for complex network
  research.}
\newblock \emph{InterJournal}, \textbf{Complex Systems}, 1695.
\newblock \urlprefix\url{http://igraph.org}.

\bibitem[{Fan \emph{et~al.}(2020)Fan, Mao, Wu, and Xu}]{GRAMPA}
Fan Z, Mao C, Wu Y, Xu J (2020).
\newblock \enquote{Spectral Graph Matching and Regularized Quadratic
  Relaxations: Algorithm and Theory.}
\newblock In HD~III, A~Singh (eds.), \emph{Proceedings of the 37th
  International Conference on Machine Learning}, volume 119 of
  \emph{Proceedings of Machine Learning Research}, pp. 2985--2995. PMLR.
\newblock \urlprefix\url{https://proceedings.mlr.press/v119/fan20a.html}.

\bibitem[{Fang \emph{et~al.}(2018)Fang, Sussman, and Lyzinski}]{soft_seeding}
Fang F, Sussman D, Lyzinski V (2018).
\newblock \enquote{Tractable Graph Matching via Soft Seeding.}

\bibitem[{Finke \emph{et~al.}(1987)Finke, Burkard, and Rendl}]{QAP}
Finke G, Burkard RE, Rendl F (1987).
\newblock \enquote{Quadratic Assignment Problems.}
\newblock In S~Martello, G~Laporte, M~Minoux, C~Ribeiro (eds.), \emph{Surveys
  in Combinatorial Optimization}, volume 132 of \emph{North-Holland Mathematics
  Studies}, pp. 61 -- 82. North-Holland.
\newblock \doi{https://doi.org/10.1016/S0304-0208(08)73232-8}.
\newblock
  \urlprefix\url{http://www.sciencedirect.com/science/article/pii/S0304020808732328}.

\bibitem[{Frank and Wolfe(1956)}]{FW}
Frank M, Wolfe P (1956).
\newblock \enquote{An algorithm for quadratic programming.}
\newblock \emph{Naval Research Logistics Quarterly}, \textbf{3}(1‐2),
  95--110.
\newblock
  \urlprefix\url{https://EconPapers.repec.org/RePEc:wly:navlog:v:3:y:1956:i:1-2:p:95-110}.

\bibitem[{Holland \emph{et~al.}(1983)Holland, Laskey, and Leinhardt}]{SBM}
Holland P, Laskey K, Leinhardt S (1983).
\newblock \enquote{Stochastic Blockmodels: First Steps.}
\newblock \emph{Social Networks - SOC NETWORKS}, \textbf{5}, 109--137.
\newblock \doi{10.1016/0378-8733(83)90021-7}.

\bibitem[{Hu \emph{et~al.}(2018)Hu, Zou, Yu, Wang, and Zhao}]{Hu2018-hd}
Hu S, Zou L, Yu JX, Wang H, Zhao D (2018).
\newblock \enquote{Answering Natural Language Questions by Subgraph Matching
  over Knowledge Graphs.}
\newblock \emph{IEEE transactions on knowledge and data engineering},
  \textbf{30}(5), 824--837.

\bibitem[{Huang \emph{et~al.}(2013)Huang, Wu, and Zhang}]{Corbi}
Huang Q, Wu LY, Zhang XS (2013).
\newblock \enquote{Corbi: a new \proglang{R} package for biological network
  alignment and querying.}

\bibitem[{Ito \emph{et~al.}(2001)Ito, Chiba, Ozawa, Yoshida, Hattori, and
  Sakaki}]{bio2}
Ito T, Chiba T, Ozawa R, Yoshida M, Hattori M, Sakaki Y (2001).
\newblock \enquote{A comprehensive two-hybrid analysis to explore the yeast
  protein interactome.}
\newblock \emph{Proceedings of the National Academy of Sciences},
  \textbf{98}(8), 4569--4574.
\newblock ISSN 0027-8424.
\newblock \doi{10.1073/pnas.061034498}.
\newblock \eprint{https://www.pnas.org/content/98/8/4569.full.pdf},
  \urlprefix\url{https://www.pnas.org/content/98/8/4569}.

\bibitem[{Jonker and Volgenant(1988)}]{lap_solver}
Jonker R, Volgenant T (1988).
\newblock \enquote{A shortest augmenting path algorithm for dense and sparse
  linear assignment problems.}
\newblock In H~Schellhaas, P~van Beek, H~Isermann, R~Schmidt, M~Zijlstra
  (eds.), \emph{DGOR/NSOR}, pp. 622--622. Springer-Verlag Berlin Heidelberg,
  Berlin, Heidelberg.
\newblock ISBN 978-3-642-73778-7.

\bibitem[{Kazemi \emph{et~al.}(2015)Kazemi, Hamed~Hassani, and
  Grossglauser}]{ExpandWhenStuck}
Kazemi E, Hamed~Hassani S, Grossglauser M (2015).
\newblock \enquote{Growing a Graph Matching from a Handful of Seeds.}
\newblock \emph{Proceedings of the VLDB Endowment}, \textbf{8}, 1010--1021.
\newblock \doi{10.14778/2794367.2794371}.

\bibitem[{Kelley \emph{et~al.}(2004)Kelley, Yuan, Lewitter, Sharan, Stockwell,
  and Ideker}]{BLAST}
Kelley B, Yuan B, Lewitter F, Sharan R, Stockwell B, Ideker T (2004).
\newblock \enquote{PathBLAST: a tool for alignment of protein interaction
  networks.}
\newblock \emph{Nucleic acids research}, \textbf{32}, W83--8.
\newblock \doi{10.1093/nar/gkh411}.

\bibitem[{Kivelä \emph{et~al.}(2014)Kivelä, Arenas, Barthelemy, Gleeson,
  Moreno, and Porter}]{multilayer}
Kivelä M, Arenas A, Barthelemy M, Gleeson JP, Moreno Y, Porter MA (2014).
\newblock \enquote{{Multilayer networks}.}
\newblock \emph{Journal of Complex Networks}, \textbf{2}(3), 203--271.
\newblock ISSN 2051-1310.
\newblock \doi{10.1093/comnet/cnu016}.
\newblock
  \eprint{https://academic.oup.com/comnet/article-pdf/2/3/203/9130906/cnu016.pdf},
  \urlprefix\url{https://doi.org/10.1093/comnet/cnu016}.

\bibitem[{Kuchaiev and Przulj(2011)}]{measure}
Kuchaiev O, Przulj N (2011).
\newblock \enquote{Integrative Network Alignment Reveals Large Regions of
  Global Network Similarity in Yeast and Human.}
\newblock \emph{Bioinformatics (Oxford, England)}, \textbf{27}, 1390--6.
\newblock \doi{10.1093/bioinformatics/btr127}.

\bibitem[{Kuhn(1955)}]{Hungarian}
Kuhn H (1955).
\newblock \enquote{The Hungarian Method for the Assignment Problem.}
\newblock \emph{Naval Res. Logist. Quart.}, \textbf{2}, 83--98.
\newblock \doi{10.1002/nav.20053}.

\bibitem[{Leskovec \emph{et~al.}(2008)Leskovec, Lang, Dasgupta, and
  Mahoney}]{Enron}
Leskovec J, Lang KJ, Dasgupta A, Mahoney MW (2008).
\newblock \enquote{Community Structure in Large Networks: Natural Cluster Sizes
  and the Absence of Large Well-Defined Clusters.}
\newblock \eprint{0810.1355}.

\bibitem[{Li and Sussman(2019)}]{Lin}
Li L, Sussman DL (2019).
\newblock \enquote{Graph Matching via Multi-Scale Heat Diffusion.}
\newblock In \emph{2019 IEEE International Conference on Big Data (Big Data)},
  pp. 1157--1162.
\newblock \doi{10.1109/BigData47090.2019.9005526}.

\bibitem[{Liu and Qiao(2012)}]{ML1}
Liu ZY, Qiao H (2012).
\newblock \enquote{A Convex-Concave Relaxation Procedure Based Subgraph
  Matching Algorithm.}
\newblock \textbf{25}, 237--252.
\newblock \urlprefix\url{http://proceedings.mlr.press/v25/liu12a.html}.

\bibitem[{{Lyzinski} \emph{et~al.}(2016){Lyzinski}, {Fishkind}, {Fiori},
  {Vogelstein}, {Priebe}, and {Sapiro}}]{FAQ}
{Lyzinski} V, {Fishkind} DE, {Fiori} M, {Vogelstein} JT, {Priebe} CE, {Sapiro}
  G (2016).
\newblock \enquote{Graph Matching: Relax at Your Own Risk.}
\newblock \emph{IEEE Transactions on Pattern Analysis and Machine
  Intelligence}, \textbf{38}(1), 60--73.
\newblock ISSN 0162-8828.
\newblock \doi{10.1109/TPAMI.2015.2424894}.

\bibitem[{Lyzinski \emph{et~al.}(2014)Lyzinski, Fishkind, and Priebe}]{SGM}
Lyzinski V, Fishkind DE, Priebe CE (2014).
\newblock \enquote{Seeded Graph Matching for Correlated Erd\"{o}s-R{\'e}nyi
  Graphs.}
\newblock \emph{J. Mach. Learn. Res.}, \textbf{15}(1), 3513--3540.
\newblock ISSN 1532-4435.
\newblock \urlprefix\url{http://dl.acm.org/citation.cfm?id=2627435.2750357}.

\bibitem[{Lyzinski and Sussman(2017)}]{row_perm}
Lyzinski V, Sussman D (2017).
\newblock \enquote{Graph Matching the Matchable Nodes when some Nodes are
  Unmatchable.}
\newblock SIAM Workshop on Network Science.

\bibitem[{Mateus \emph{et~al.}(2008)Mateus, Horaud, Knossow, Cuzzolin, and
  Boyer}]{SpecMatch}
Mateus D, Horaud RP, Knossow D, Cuzzolin F, Boyer E (2008).
\newblock \enquote{Articulated Shape Matching Using Laplacian Eigenfunctions
  and Unsupervised Point Registration.}
\newblock In \emph{Proceedings of the IEEE Conference on Computer Vision and
  Pattern Recognition}.
\newblock
  \urlprefix\url{http://perception.inrialpes.fr/Publications/2008/MHKCB08}.

\bibitem[{Moorman \emph{et~al.}(2018)Moorman, Chen, Tu, Boyd, and
  Bertozzi}]{Moorman2018-ha}
Moorman JD, Chen Q, Tu TK, Boyd ZM, Bertozzi AL (2018).
\newblock \enquote{Filtering Methods for Subgraph Matching on Multiplex
  Networks.}
\newblock In \emph{2018 {IEEE} International Conference on Big Data (Big
  Data)}, pp. 3980--3985. ieeexplore.ieee.org.

\bibitem[{Nabieva \emph{et~al.}(2005)Nabieva, Jim, Agarwal, Chazelle, and
  Singh}]{bio3}
Nabieva E, Jim K, Agarwal A, Chazelle B, Singh M (2005).
\newblock \enquote{Whole-proteome prediction of protein function via
  graph-theoretic analysis of interaction maps.}

\bibitem[{{Narayanan} and {Shmatikov}(2009)}]{SocialNetwork}
{Narayanan} A, {Shmatikov} V (2009).
\newblock \enquote{De-anonymizing Social Networks.}
\newblock In \emph{2009 30th IEEE Symposium on Security and Privacy}, pp.
  173--187.
\newblock \doi{10.1109/SP.2009.22}.

\bibitem[{Papadimitriou and Steiglitz(1998)}]{Papadimitriou1998-lz}
Papadimitriou CH, Steiglitz K (1998).
\newblock \emph{Combinatorial Optimization: Algorithms and Complexity}.
\newblock Courier Corporation.
\newblock
  \urlprefix\url{https://play.google.com/store/books/details?id=cDY-joeCGoIC}.

\bibitem[{Riccardo and Marc(2015)}]{transportation}
Riccardo G, Marc B (2015).
\newblock \enquote{The Multilayer Temporal Network of Public Transport in Great
  Britain.}
\newblock \emph{Scientific Data}, \textbf{2}.
\newblock \doi{10.1038/sdata.2014.56}.

\bibitem[{Singh \emph{et~al.}(2008)Singh, Xu, and Berger}]{IsoRank}
Singh R, Xu J, Berger B (2008).
\newblock \enquote{Global alignment of multiple protein interaction networks
  with application to functional orthology detection.}
\newblock \emph{Proceedings of the National Academy of Sciences},
  \textbf{105}(35), 12763--12768.
\newblock ISSN 0027-8424.
\newblock \doi{10.1073/pnas.0806627105}.
\newblock \eprint{https://www.pnas.org/content/105/35/12763.full.pdf},
  \urlprefix\url{https://www.pnas.org/content/105/35/12763}.

\bibitem[{{Sussman} \emph{et~al.}(2018){Sussman}, {Lyzinski}, {Park}, and
  {Priebe}}]{centering}
{Sussman} DL, {Lyzinski} V, {Park} Y, {Priebe} CE (2018).
\newblock \enquote{{Matched Filters for Noisy Induced Subgraph Detection}.}
\newblock \emph{arXiv e-prints}, arXiv:1803.02423.
\newblock \eprint{1803.02423}.

\bibitem[{{Umeyama}(1988)}]{Umeyama}
{Umeyama} S (1988).
\newblock \enquote{An eigendecomposition approach to weighted graph matching
  problems.}
\newblock \emph{IEEE Transactions on Pattern Analysis and Machine
  Intelligence}, \textbf{10}(5), 695--703.
\newblock ISSN 0162-8828.
\newblock \doi{10.1109/34.6778}.

\bibitem[{Volgenant(1996)}]{Volgenant1996-o}
Volgenant A (1996).
\newblock \enquote{Linear and semi-assignment problems: A core oriented
  approach.}
\newblock \emph{Computer and Operations Research}, \textbf{23}, 917--932.

\bibitem[{Yartseva and Grossglauser(2013)}]{Percolation}
Yartseva L, Grossglauser M (2013).
\newblock \enquote{On the Performance of Percolation Graph Matching.}
\newblock pp. 119--130.
\newblock \doi{10.1145/2512938.2512952}.
\newblock \urlprefix\url{http://doi.acm.org/10.1145/2512938.2512952}.

\bibitem[{Young and Scheinerman(2007)}]{rdpg}
Young SJ, Scheinerman ER (2007).
\newblock \enquote{Random Dot Product Graph Models for Social Networks.}
\newblock In \emph{Proceedings of the 5th International Conference on
  Algorithms and Models for the Web-graph}, WAW'07, pp. 138--149.
  Springer-Verlag, Berlin, Heidelberg.
\newblock ISBN 3-540-77003-8, 978-3-540-77003-9.
\newblock \urlprefix\url{http://dl.acm.org/citation.cfm?id=1777879.1777890}.

\bibitem[{{Zaslavskiy} \emph{et~al.}(2009){Zaslavskiy}, {Bach}, and
  {Vert}}]{PATH}
{Zaslavskiy} M, {Bach} F, {Vert} J (2009).
\newblock \enquote{A Path Following Algorithm for the Graph Matching Problem.}
\newblock \emph{IEEE Transactions on Pattern Analysis and Machine
  Intelligence}, \textbf{31}(12), 2227--2242.
\newblock ISSN 0162-8828.
\newblock \doi{10.1109/TPAMI.2008.245}.

\end{thebibliography}

\end{document}